\documentclass[12pt]{article}
\pdfoutput=1
\usepackage{a4wide,epsfig,psfrag,amsmath,amssymb,cite,scalefnt}
\usepackage{color}
\usepackage{amsmath,comment,braket}
\usepackage{placeins}
\usepackage{breqn}

\graphicspath{ {figs/} }

\allowdisplaybreaks

\begin{document}


\title{\vskip-3cm{\baselineskip14pt
    \begin{flushleft}
      \normalsize 
      CERN-TH-2019-103\\
      MPP-2019-128\\
      P3H-19-015\\
      TTP19-018\\
      ZU-TH 34/19
  \end{flushleft}}
  \vskip1.5cm
  Double Higgs boson production at NLO: combining the exact numerical result and high-energy expansion
}

\author{
  Joshua Davies$^{a}$,
  Gudrun Heinrich$^{b}$,
  Stephen P. Jones$^{c}$,
  Matthias Kerner$^{d}$,\\
  Go Mishima$^{a,e}$,
  Matthias Steinhauser$^{a}$,
  David Wellmann$^{a}$
  \\[1mm]
  {\small\it $^a$Institut f{\"u}r Theoretische Teilchenphysik}\\
  {\small\it Karlsruhe Institute of Technology (KIT)}\\
  {\small\it Wolfgang-Gaede Stra\ss{}e 1, 76128 Karlsruhe, Germany}
  \\[1mm]
  {\small\it $^b$Max Planck Institute for Physics}\\
  {\small\it F\"ohringer Ring 6, 80805 M\"unchen, Germany}
  \\[1mm]
  {\small\it $^c$Theoretical Physics Department, CERN}\\
  {\small\it Geneva, Switzerland}
  \\[1mm]
  {\small\it $^d$Physik-Institut, Universit\"at Z\"urich,}\\
  {\small\it Winterthurerstrasse 190, 8057 Z\"urich, Switzerland}
  \\[1mm]
  {\small\it $^e$Institut f{\"u}r Kernphysik}\\
  {\small\it Karlsruhe Institute of Technology (KIT)}
  \\
  {\small\it Hermann-von-Helmholtz-Platz 1, 76344 Eggenstein-Leopoldshafen, Germany}
}
  
\date{}

\maketitle

\thispagestyle{empty}

\begin{abstract}
  We consider the next-to-leading order QCD corrections to Higgs boson pair production,
  using our recent calculation of the form factors in the high-energy limit.
  We compute the virtual corrections to the partonic cross
  section, applying Pad\'e approximations to extend the
  range of validity of the high-energy expansion. This enables us to compare
  to the exact numerical calculation in a significant part of the phase space
  and allows us to extend the virtual matrix element grid, based on the exact numerical
  calculation, to larger values of the (partonic) transverse momentum of the
  Higgs boson, which is important for boosted Higgs studies. Improved predictions 
  for hadron colliders with centre-of-mass energies of $14\ \mathrm{TeV}$ and 
  $100\ \mathrm{TeV}$ are presented. The updated grid is made publicly available.
\end{abstract}

\thispagestyle{empty}

\sloppy


\newpage


\section{Introduction}
\label{sec:intro}

\newcommand{\mhh}{m_{hh}}
\newcommand{\ftapprox}{FT$_{\small\mbox{approx}}$}

A primary goal of the Large Hadron Collider (LHC) and future colliders is the exploration of the
electroweak symmetry breaking sector of the Standard Model (SM).
It is important to find out whether the Higgs potential is indeed of the form
suggested by the Standard Model, where the trilinear and quartic Higgs boson
self-couplings are completely determined by the Higgs boson mass and its
vacuum expectation value.  As a deviation of the trilinear coupling from its
SM value would be a clear sign of physics beyond the Standard Model, it is
crucial to have precise predictions for processes which allow the measurement
of this coupling.

An important process in this regard is Higgs boson pair production via gluon
fusion. While the cross section is about a factor of one thousand smaller than that
of single Higgs boson production, it is nevertheless the Higgs boson pair production
channel with the largest cross section.  It
also has the interesting feature that there is a delicate cancellation between
triangle-type diagrams, containing the trilinear Higgs boson coupling
$\lambda$, and box-type diagrams, containing only Yukawa couplings, such that
deviations of the trilinear coupling from the SM value can lead to distinct
features in observables such as the Higgs boson pair invariant mass ($\mhh$)
distribution.

Measurements of double Higgs boson production in gluon fusion at the LHC already
have led to constraints on the ratio
$\kappa_\lambda=\lambda_{\rm{BSM}}/\lambda_{\rm{SM}}$~\cite{Sirunyan:2018two,Sirunyan:2018iwt,ATLAS:2018otd,Aaboud:2018ftw},
where currently $-5.0\leq \kappa_\lambda\leq 12.0$ at 95\% confidence
level~\cite{ATLAS:2018otd} is the most stringent bound derived from
Higgs boson pair production measurements.  The tightest bounds
typically result from the combination of various Higgs boson decay channels.
Among these, an important channel is the $b\bar{b}b\bar{b}$
channel~\cite{Sirunyan:2018tki,Aaboud:2018knk} due to the large branching
ratio of $H\to b\bar{b}$.  Reconstructing the Higgs bosons from boosted jets
is promising, not only in view of a potential 27\,TeV or 100\,TeV collider,
but also at the HL-LHC~\cite{deLima:2014dta}.  However, such an analysis also
requires precise predictions in the high-$p_T$, or large-$\mhh$ regime of the
Higgs bosons, reaching $\mhh$ values of about 3\,TeV at the LHC, which is a
region where high-energy expansions are fully justified.  In this paper we
will combine the high energy expansion of
Refs.~\cite{Davies:2018ood,Davies:2018qvx,Mishima:2018olh} with the full NLO
calculation of Refs.~\cite{Borowka:2016ehy,Borowka:2016ypz,Heinrich:2017kxx}
to arrive at predictions which combine the virtues of both approaches in the
kinematic ranges where they work best.

The leading-order (LO) contribution to Higgs boson pair production in gluon fusion 
already contains one loop, therefore the calculation of higher-order
corrections is a complex task. While the LO calculation was
performed some time
ago~\cite{Eboli:1987dy,Glover:1987nx,Plehn:1996wb}, next-to-leading order (NLO) 
results with full top quark mass dependence became available only
recently~\cite{Borowka:2016ehy,Borowka:2016ypz,Baglio:2018lrj}, based on a
numerical evaluation of the corresponding two-loop integrals.

Analytic higher-order results are known in various approximations.  NLO QCD
corrections in the $m_t\to\infty$ limit, sometimes also called ``Heavy Top
Limit (HTL)'', or ``Higgs Effective Field Theory~(HEFT)'' approximation, have
been calculated in Ref.~\cite{Dawson:1998py} using the so-called
``Born-improved HTL'' approximation, which involves rescaling the NLO results
in the $m_t\to\infty$ limit by a factor $B_{\rm FT}/B_{\rm HTL}$, where
$B_{\rm FT}$ denotes the squared LO matrix element in the full theory.  In
Ref.~\cite{Maltoni:2014eza} an approximation called ``\ftapprox'' was
introduced, which contains the real radiation matrix elements with full top
quark mass dependence, while the virtual part is calculated in the
Born-improved HTL approximation.

The next-to-next-to-leading order (NNLO) QCD corrections in the $m_t\to\infty$
limit have been computed in
Refs.~\cite{deFlorian:2013uza,deFlorian:2013jea,Grigo:2014jma,deFlorian:2016uhr}.
The HTL results have also been improved in various ways: the virtual
corrections have been supplemented by an expansion in $1/m_t^2$
in~\cite{Grigo:2013rya,Grigo:2015dia} up to order $1/m_t^{12}$ at NLO and
$1/m_t^{4}$ at NNLO. Real radiation corrections, which involve three closed
top quark loops have been expanded up to $1/m_t^8$ in
Ref.~\cite{Davies:2019xzc}.  Soft gluon resummation has been performed at
NNLO+NNLL level in~\cite{deFlorian:2015moa}.  In Ref.~\cite{Grazzini:2018bsd},
the NNLO calculation in the HTL of Ref.~\cite{deFlorian:2016uhr} has been
combined with results including the full top quark mass dependence at NLO as
well as in the $2\to 4$ matrix elements present in the NNLO real
radiation. The latter results have been supplemented by soft gluon resummation
in Ref.~\cite{deFlorian:2018tah}.  Analytic approximations for the top quark
mass dependence of the two-loop amplitudes entering $gg\to HH$ at NLO have
also been studied in the high-energy
limit~\cite{Davies:2018ood,Davies:2018qvx,Mishima:2018olh}, around the top
pair threshold expansion combined with large mass
expansion~\cite{Grober:2017uho}, and for small Higgs boson transverse
momentum~\cite{Bonciani:2018omm}.

The full NLO calculation of Refs.~\cite{Borowka:2016ehy,Borowka:2016ypz} has
been combined~\cite{Heinrich:2017kxx,Jones:2017giv,Heinrich:2019bkc} with
parton showers within the {\tt
  POWHEG-BOX-V2}~\cite{Nason:2004rx,Frixione:2007vw,Alioli:2010xd} and {\tt
  MG5\_aMC@NLO}~\cite{Alwall:2014hca,Hirschi:2015iia} frameworks as well as
within Sherpa~\cite{Gleisberg:2008ta}. Ref.~\cite{Heinrich:2019bkc} contains a
discussion of showered results for non-SM values of the trilinear Higgs
coupling, as well as a comparison of {\tt Pythia8.2}~\cite{Sjostrand:2014zea}
and {\tt Herwig7.1}~\cite{Bellm:2017bvx} showers in combination with {\tt
  Powheg}.

The purpose of this paper is to provide results for the process $gg\to HH$ at
NLO which are valid and accurate in the low-, medium- and high-energy regimes.  This is achieved
by combining the high-energy expansion, computed in
Refs.~\cite{Davies:2018ood,Davies:2018qvx,Mishima:2018olh}, with the existing grid of the exact NLO
result~\cite{Heinrich:2017kxx,hhgrid},
such that the finite part of the virtual amplitude can be evaluated at any phase space
point without having to do costly two-loop numerical integrations. Previously, the grid of the exact NLO
result was constructed based only on unweighted events, which are sparse in the high-energy
region, and the grid was therefore statistically limited in the high-energy region.
Extending the grid to higher energies using the exact NLO result would require the costly evaluation of 
additional phase-space points in a regime where the numerical convergence of the two-loop integrals can be poor.
Instead, by combining the existing grid with analytic results obtained through a high-energy
expansion, after a careful assessment of the regions in which the latter leads to
an improvement, we are able to present results with small uncertainties over the full kinematic range.  
This improvement is particularly relevant for highly boosted Higgs bosons, for which the previous grid was unreliable. 
Parton shower Monte Carlo programs based on the new grid, presented here, can reliably be used 
to make predictions in an extended kinematic range.

The remainder of the paper is structured as follows. In the next section we
introduce our notation and in Section~\ref{sec::virt} we describe our
approach to obtain Pad\'e approximations for the NLO virtual corrections based on the
high-energy expansion of the form factors. This approach is validated in
Section~\ref{sec::master} at the level of the master integrals. In
Section~\ref{sec::Vfin} we present numerical results for the virtual
corrections and, in Section~\ref{sec::appl}, we study their impact on the transverse momentum and invariant mass
distributions. We conclude in
Section~\ref{sec::concl}.


\section{Notation and conventions}

The analysis we perform in this paper is based on the results for the form
factors obtained in Refs.~\cite{Davies:2018ood,Davies:2018qvx}.
Let us briefly repeat the notation and conventions introduced
in these references.

The amplitude for the process $g(q_1)g(q_2)\to H(q_3)H(q_4)$, with all momenta
$q_i$ defined to be incoming, can be decomposed into two Lorentz structures
\begin{eqnarray}
  {\cal M}^{ab} &=& 
  \varepsilon_{1,\mu}\varepsilon_{2,\nu}
  {\cal M}^{\mu\nu,ab}
  \,\,=\,\,
  \varepsilon_{1,\mu}\varepsilon_{2,\nu}
  \delta^{ab} X_0 s 
  \left( F_1 A_1^{\mu\nu} + F_2 A_2^{\mu\nu} \right)
  \,,
                    \label{eq::M}
\end{eqnarray}
where $a$ and $b$ are adjoint colour indices,
$s=(q_1+q_2)^2$ is the squared partonic centre-of-mass energy
 and the two Lorentz structures are given by
\begin{eqnarray}
  A_1^{\mu\nu} &=& g^{\mu\nu} - {\frac{1}{q_{12}}q_1^\nu q_2^\mu
  }\,,\nonumber\\
  A_2^{\mu\nu} &=& g^{\mu\nu}
                   + \frac{1}{{p_T^2} q_{12}}\left(
                   q_{33}    q_1^\nu q_2^\mu
                   - 2q_{23} q_1^\nu q_3^\mu
                   - 2q_{13} q_3^\nu q_2^\mu
                   + 2q_{12} q_3^\mu q_3^\nu \right)\,,
\end{eqnarray}
with
\begin{eqnarray}
  q_{ij} &=& q_i\cdot q_j\,,\qquad
  {p_T^{\:2}} \:\:\:=\:\:\: \frac{2q_{13}q_{23}}{q_{12}}-q_{33}
  {\:\:\:=\:\:\: \frac{tu - m_h^4}{s}}
  \,,\nonumber\\
  X_0 &=& \frac{G_F}{\sqrt{2}} \frac{\alpha_s(\mu)}{2\pi} T_F \,,
\end{eqnarray}
where $s,{t=(q_1+q_3)^2}$ and ${u=(q_2+q_3)^2}$ are Mandelstam variables which fulfill
$s+t+u=2m_h^2$, $T_F=1/2$, $G_F$ is Fermi's constant and $\alpha_s(\mu)$ is the strong
coupling constant evaluated at the renormalization scale $\mu$.

\begin{figure}[t]
  \centering
  \begin{tabular}{ccc}
    \includegraphics[width=.3\textwidth]{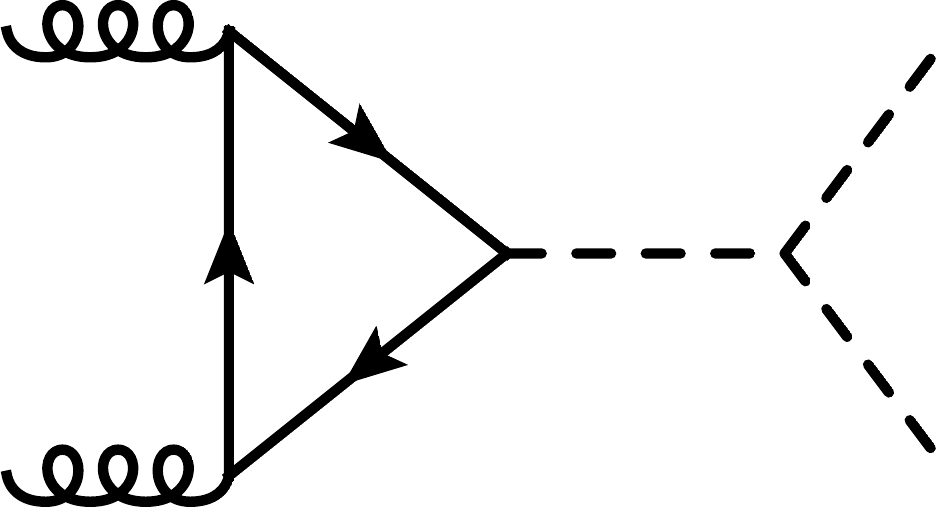} &
    \includegraphics[width=.3\textwidth]{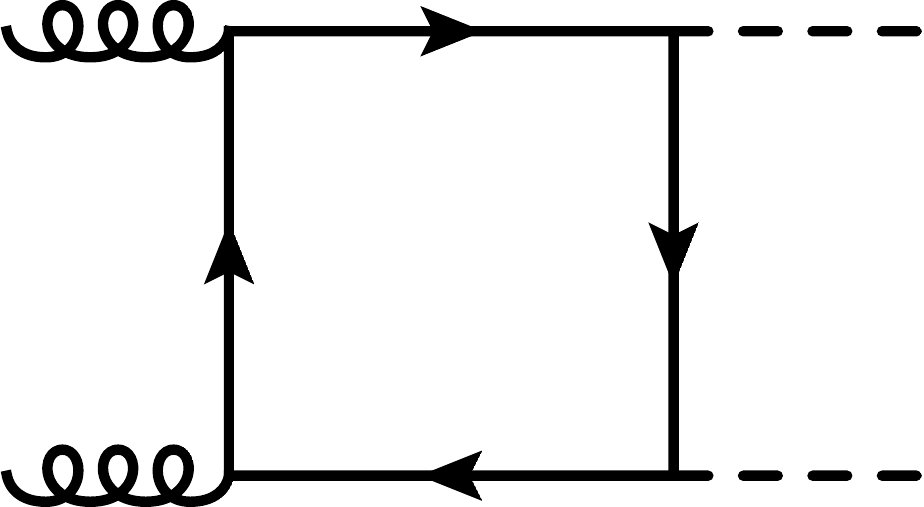} &
    \includegraphics[width=.3\textwidth]{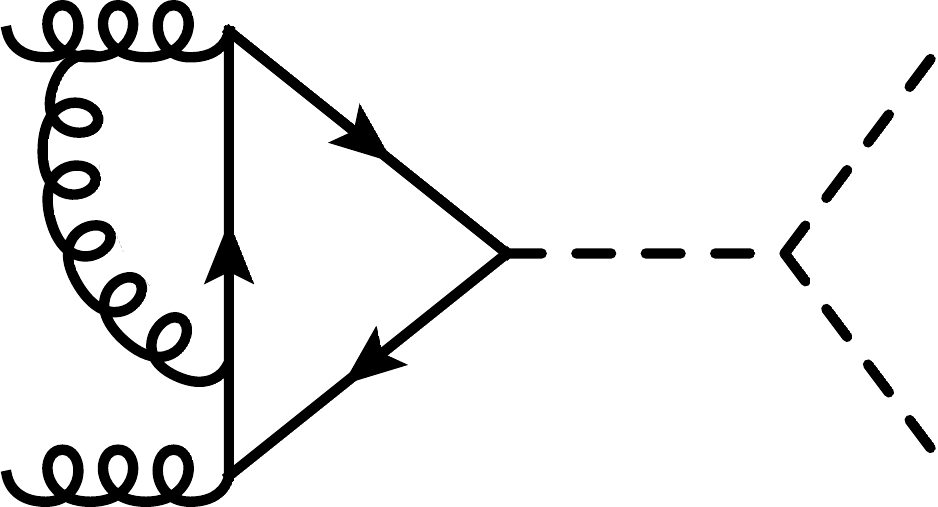} 
    \\
    (a) & (b) & (c) \\
    \raisebox{1.em}{\includegraphics[width=.3\textwidth]{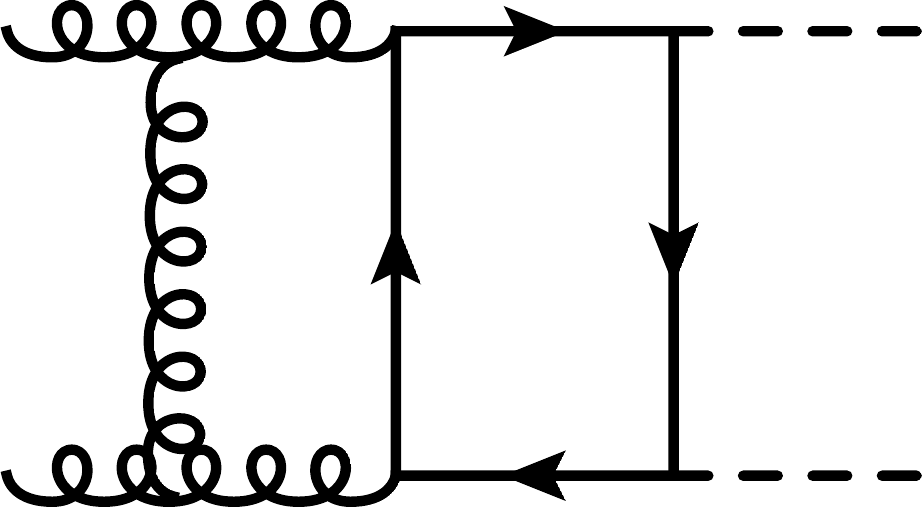}} &
    \raisebox{1.em}{\includegraphics[width=.3\textwidth]{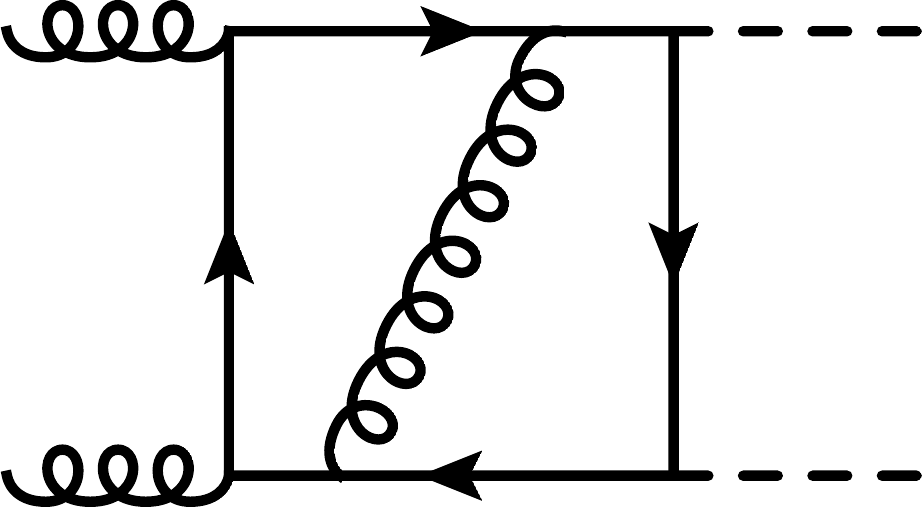}} &
    \includegraphics[width=.23\textwidth]{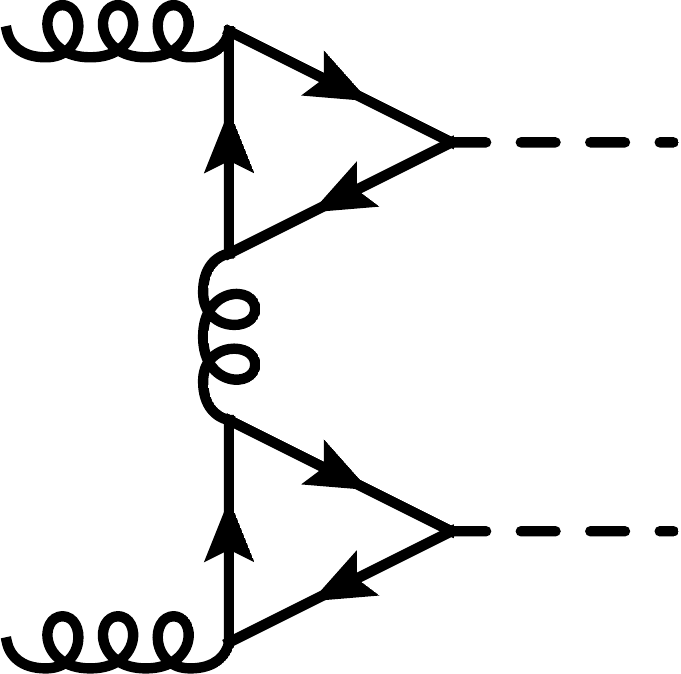} 
    \\
    (d) & (e) & (f)
  \end{tabular}
  \caption{\label{fig::dia}
    One- and two-loop Feynman diagrams contributing to $gg\to HH$.
    Solid, curly, and dashed lines represent fermions, gluons, and Higgs bosons respectively.
    }
\end{figure}

We define the expansion in $\alpha_s$ of the form factors as
\begin{eqnarray}
  F &=& F^{(0)} + \frac{\alpha_s(\mu)}{\pi} F^{(1)} + \cdots
  \,,
  \label{eq::F}
\end{eqnarray}
and decompose the functions $F_1$ and $F_2$ introduced in Eq.~(\ref{eq::M})
into ``triangle'' and ``box'' form factors. We thus cast the one- and two-loop
corrections in the form
\begin{eqnarray}
  F_1^{(0)} &=& \frac{3 m_h^2}{s-m_h^2} F^{(0)}_{\rm tri}+F^{(0)}_{\rm box1}\,,
  \nonumber\\
  F_2^{(0)} &=& F^{(0)}_{\rm box2}\,, \nonumber\\
  F_1^{(1)} &=& \frac{3 m_h^2}{s-m_h^2} F^{(1)}_{\rm tri}+F^{(1)}_{\rm box1}
                +F^{(1)}_{\rm dt1}\,, \nonumber\\
  F_2^{(1)} &=& F^{(1)}_{\rm box2}+F^{(1)}_{\rm dt2}\,.
                \label{eq::F_12}
\end{eqnarray}
$F^{(1)}_{\rm dt1}$ and $F^{(1)}_{\rm dt2}$ denote the contribution
from one-particle reducible diagrams such as the one shown in Fig.~\ref{fig::dia}(f).
In Ref.~\cite{Davies:2018qvx} this contribution has not been considered since
the full top quark mass dependence is available from Eqs.~(24),~(25) and~(26)
of Ref.~\cite{Degrassi:2016vss}.

At this point a comment on the definition of $\alpha_s$ is in order.
In Ref.~\cite{Davies:2018qvx} $\alpha_s$ has been defined with six active
flavours which is an appropriate choice for the high-energy limit.
In this paper, we compare to Ref.~\cite{Borowka:2016ypz}
where a five-flavour $\alpha_s$ has been used. 
Thus, we have to transform $\alpha_s$ and the gluon wave function from the 
six-flavour to the five-flavour theory using the relations
\begin{eqnarray}
  \alpha_s^{(6)}(\mu) &=& \alpha_s^{(5)}(\mu)\left( 1 + \frac{\alpha_s^{(5)}(\mu)}{3\pi}T_F\log\frac{\mu^2}{m_t^2} + \mathcal{O}(\alpha_s^2) \right)
                          \,,
  \\
  A_\nu^{(6)}(\mu) &=& A_\nu^{(5)}(\mu)\left(1 - \frac{\alpha_s^{(5)}(\mu)}{3\pi}T_F\log\frac{\mu^2}{m_t^2} + \mathcal{O}(\alpha_s^2) \right)
                   \,,
\end{eqnarray}
where $A_\nu$ is the gluon wave function.
As can be seen from these expressions the additional terms cancel
because the number of external gluon fields equals the number of
strong couplings $g_s$ in the Born amplitude, such that  the resulting analytic
expressions remain identical.

After the renormalization of the ultraviolet divergences, the form factors are
still infrared divergent. Infrared finite results can be obtained
by making a suitable subtraction~\cite{Catani:1998bh}
\begin{eqnarray}
  F^{{\rm fin},(1)} = F^{(1),\rm IR} - K_g^{(1)} F^{(0)},
\end{eqnarray}
where $K_g^{(1)}$ is given by
\begin{eqnarray}
  K_g^{(1)} &=& - \left(\frac{\mu^2}{-s-i\delta}\right)^\epsilon 
  \frac{e^{\epsilon\gamma_E}}{ 2 \Gamma(1-\epsilon)}
  \left[\frac{C_A}{\epsilon^2} +
    \frac{1}{\epsilon}\left( \frac{11}{6}C_A-\frac{2}{3}T_Fn_l \right)
  \right]
  \,.
  \label{eq::Ig1}
\end{eqnarray}
$\gamma_E$ is Euler's constant and $C_A=3$ is a colour factor.
After the decoupling of the top quark we have $n_l=5$ as the
number of active flavours.\footnote{In Ref.~\cite{Davies:2018qvx}
  infrared subtraction has been performed in QCD with six active flavours.}
Note that the choice of $K_g^{(1)}$ is not unique. For example, finite form
factors are also obtained if the $\mu$-dependent factor multiplies only
the $1/\epsilon^2$ term inside the square brackets of Eq.~(\ref{eq::Ig1}),
and not the $1/\epsilon$ term.
The resulting form factors differ by terms proportional to
$\log(\mu^2/(-s-i\delta))$.  For the definition of $K_g^{(1)}$ in Eq.~(\ref{eq::Ig1}) it is
convenient to introduce
\begin{eqnarray}
  F^{{\rm fin},(1)} &=& \tilde{F}^{(1)}
              + \beta_0 \tilde{F}^{(0)} \log\left(\frac{\mu^2}{-s-i\delta}\right)
              \,,
\end{eqnarray}
where $\tilde{F}^{(i)} = F^{{\rm fin},(i)}(\mu^2=-s)$,
and $\beta_0=11 C_A/12 - T_F n_l/3$.

In Ref.~\cite{Davies:2018qvx} we express the analytic results for the form
factors in terms of $m_t, s, t$ and $m_h$.  Note that our two-loop expressions
are Taylor expanded\footnote{Let us stress that only the form factors are
  expanded in $m_h$ and the factor $3 m_h^2/(s-m_h^2)$ in Eq.~(\ref{eq::F_12}) is
  kept exact.} in $m_h$ including terms up to $m_h^2$.  In
Section~\ref{sec::virt} we use the (infrared-finite) form factors to construct
the virtual corrections for the process $gg\to HH$. We adapt the notation of
Ref.~\cite{Borowka:2016ypz} and express our results in terms of the variables
$m_t, s, m_h$ and $p_T^2$. This is achieved using the relation
\begin{eqnarray}
  t &=& m_h^2 - \frac{s}{2}\left(1 - \sqrt{1-4\,\frac{m_h^2+p_T^2}{s}} \right)
        \,,
        \label{eq::t}
\end{eqnarray}
and a subsequent re-expansion of the two-loop form factors in $m_h$ up to order
$m_h^2$. We use the exact expression for the one-loop
corrections~\cite{Glover:1987nx,Plehn:1996wb} and thus no expansion is
necessary.


\section{\label{sec::virt}Pad\'e improved virtual corrections}

We adapt the notation of Ref.~\cite{Heinrich:2017kxx} and define (see also Eq.~(4.1)
of Ref.~\cite{Grober:2017uho})
\begin{eqnarray}
  \widetilde{\mathcal{V}}_{\textnormal{fin}} &=&
  \frac{\alpha_s^2\left(\mu\right)}{16\pi^2}\frac{G_F^2 s^2}{64}
  \left[   C + 2\left( {\tilde{F}_1^{(0)*}}
    \tilde{F}_1^{(1)} + {\tilde{F}_2^{(0)*}} \tilde{F}_2^{(1)} +
    \tilde{F}_1^{(0)} {\tilde{F}_1^{(1)*}}+\tilde{F}_2^{(0)}
          {\tilde{F}_2^{(1)*}} \right) \right] \,,
      \label{eq::Vtil}
\end{eqnarray}
with
\begin{eqnarray}
  C &=& \left[\left|\tilde{F}_1^{(0)}\right|^2+\left|\tilde{F}_2^{(0)}\right|^2\right]
        \left(
        C_A\pi^2 - C_A\log^2\frac{\mu^2}{s}
        \right)
        \,.
\end{eqnarray}
Here $\alpha_s$ corresponds to the five-flavour strong coupling constant. 
Furthermore, we introduce
\begin{eqnarray}
  \mathcal{V}_{\textnormal{fin}} 
  &=& \frac{\widetilde{\mathcal{V}}_{\textnormal{fin}}}{\alpha_s^2(\mu)}
      \,.
\end{eqnarray}

In Ref.~\cite{hhgrid} a grid of
3398 phase-space points is provided in the file \texttt{Virt\_full\_noas.grid}
where the result for the phase-space point $P_i=(s_i,t_i)$ is
given in the format
\begin{equation}
	\left(\; \beta(s_i)\;,\; \cos(\theta_i)\;,\; \mathcal{V}_{\rm
	fin}^{\rm grid}\left(P_i\right)\;,\; \pm \delta_i \;\right)
\end{equation}
with the coordinates $\beta(s)$ and $\cos(\theta)$ given by
(see page 4 of Ref. \cite{Heinrich:2017kxx})
\begin{equation}
	\beta(s)=\sqrt{1-\frac{4m_h^2}{s}} \qquad {\rm and} \qquad
	\cos(\theta)=\frac{s+2t-2m_h^2}{s\beta(s)} \,.
\end{equation}
We use Eq.~(\ref{eq::t}) together with
\begin{equation}
	p_T^2=\frac{tu - m_h^4}{s} \qquad {\rm and} \qquad u=2m_h^2-s-t
\end{equation}
in order to switch to the coordinates $({\sqrt{s}},p_T)$ in the following.

For the numerical evaluation of ${\cal V}_{\rm fin}$ we proceed as follows:
\begin{itemize}
\item After inserting into Eq.~(\ref{eq::Vtil}) the exact one-loop and
  two-loop one-particle reducible form factors and the
  high-energy expansion of the remaining parts ${\cal V}_{\rm fin}$ can be
  written as
  \begin{eqnarray}
    {\cal V}_{\rm fin}^N &=& \mathcal{V}_{0}+\sum_{i=2}^{N}\mathcal{V}_{i}\, m_t^i
                           \,,
                           \label{eq::Vfin_exp}
  \end{eqnarray}
  where $\mathcal{V}_{0}$ contains all parts that are exact in $m_t$ and $m_h$
  (i.e., $F^{(0)}_{\textnormal{tri}}$, $F^{(0)}_{\textnormal{box1}}$,
  $F^{(0)}_{\textnormal{box2}}$, $F^{(1)}_{\textnormal{dt1}}$ and
  $F^{(1)}_{\textnormal{dt2}}$) and the second term in
  Eq.~(\ref{eq::Vfin_exp}) contains those parts which involve\footnote{Exact
    results for $F^{(1)}_{\textnormal{tri}}$ are available from
    Refs.~\cite{Harlander:2005rq,Anastasiou:2006hc,Aglietti:2006tp}. For simplicity, in the following we
    nevertheless use our expansions which provide a very good approximation of
    the exact result~\cite{Davies:2018qvx}.} $F^{(1)}_{\textnormal{tri}}$,
  $F^{(1)}_{\textnormal{box1}}$ and $F^{(1)}_{\textnormal{box2}}$.  In
  Eq.~(\ref{eq::Vfin_exp}) we explicitly show the dependence on $m_t$ but
  suppress dependence on $m_h$; note that ${\cal V}_i$ contains an expansion up to
  $m_h^2$.

\item
  At this point we fix all numerical values except the top quark mass, 
  i.e., $\sqrt{s}$, $p_T$ and $m_h$.

\item
  Next we apply the replacements\footnote{$\log m_t$ terms
    are not replaced.} $m_t^{2k} \to m_t^{2k} x^k$
  and $m_t^{2k-1} \to m_t^{2k-1} x^k$ for the odd and even powers of $m_t$. We
  insert the numerical value for $m_t$ 
  and consider ${\cal V}_{\rm fin}^N$ as an expansion in $x$.
  In Ref.~\cite{Davies:2018qvx} terms up to order $m_t^{16}$ were
  presented. Since then the expansion has been extended to $m_t^{32}$
  which implies that ${\cal V}_{\rm fin}^N$ is available up to $x^{16}$.
  The analytic results for the form factors can be obtained from~\cite{progdata_appl}.

\item
  Next we construct Pad\'e approximants of ${\cal V}_{\rm fin}^N$ in
  the variable $x$ which means that we write Eq.~(\ref{eq::Vfin_exp})
  as a rational function of the form
  \begin{eqnarray}
    {\cal V}_{\rm fin}^N
    &=& 
        \frac{a_0 + a_1 x + \ldots + a_n x^n}{1 + b_1 x + \ldots + b_m x^m}
          \,\, \equiv\,\, [n/m](x)\,,
          \label{eq::Pade}
  \end{eqnarray}
  where $a_i$ and $b_i$ are determined by comparing the coefficients of $x^k$
  after expanding the right-hand side of Eq.~(\ref{eq::Pade}) in $x$.

  As an alternative approach one can construct
  Pad\'e approximations for ${\cal V}_{\rm fin}^N - {\cal V}_0$, which have $a_0=0$ and
  different values for the remaining coefficients. Both approaches lead to
  very similar final results, so in our
  analysis we concentrate on the one outlined in Eq.~(\ref{eq::Pade}).

\item
  For $N=32$, Pad\'e approximations with $n+m=16$ can be constructed.
  We restrict our analysis to Pad\'e approximants which are close to
  ``diagonal'' (where $n=m$). We require $|n-m|\le2$.
  Furthermore, we demand that expansions include
  at least terms up to order $m_t^{30}$. This leads to a list of five Pad\'e 
  approximants $\mathbb{Q}=\left\lbrace [7/8],[8/7],[7/9],[8/8],[9/7] \right\rbrace$.

\item We aim for an approximation of ${\cal V}_{\rm fin}$ in the
  two-dimensional $\sqrt{s}$--$p_T$ plane where for each point a separate
  Pad\'e approximant is constructed.  Due to the structure of the
  ansatz (Eq.~(\ref{eq::Pade})), the Pad\'e approximants may develop poles
  in the complex $x$ plane.  Poles close
  to $x=1$ might lead to unphysical results. For this reason we assign a
  weight to each Pad\'e approximant, which depends on the distance of the
  closest pole to $x=1$, and use this information to construct for each pair
  $(\sqrt{s},p_T)$ a central value and an estimate of the uncertainty. In
  detail, we proceed as follows
  \begin{itemize}
  \item For each phase-space point $(\sqrt{s},p_T)$
    we compute for all Pad\'e approximants in $\mathbb{Q}$ (see
    above) the value at $x=1$ and the distance of the closest pole
    which we denote by $\alpha_i$ and $\beta_i$, respectively.
  \item We introduce a re-weighting function, which reduces the impact
    of values $\alpha_i$ from Pad\'e approximations with
    poles close to $x=1$. We define
    \begin{eqnarray}
      \omega_i &=& \frac{\beta_i^2}{\sum_j\beta_j^2}\,,
    \end{eqnarray}
    and assign $\omega_i$ to each value $\alpha_i$.    
  \item We use the values $\alpha_i$ and $\omega_i$ to compute the central
    value from the weighted average and the uncertainty from the standard
    deviation as follows
    \begin{eqnarray}
      \alpha &=& \sum_i\omega_i\alpha_i \,,
      \nonumber\\
      \delta_{\alpha} &=& \sqrt{\frac{\sum_i\omega_i\left(\alpha_i-\alpha\right)^2}{1-\sum_i\omega_i^2}}
      \,.
    \end{eqnarray}
\end{itemize}
This procedure provides for each point $(\sqrt{s},p_T)$ a
result of the form $\alpha\pm \delta_{\alpha}$
which is based on Pad\'e approximation. 

\end{itemize}


\section{\label{sec::master}Pad\'e improved master integrals}

\begin{figure}[t]
  \centering
    \includegraphics[width=\textwidth]{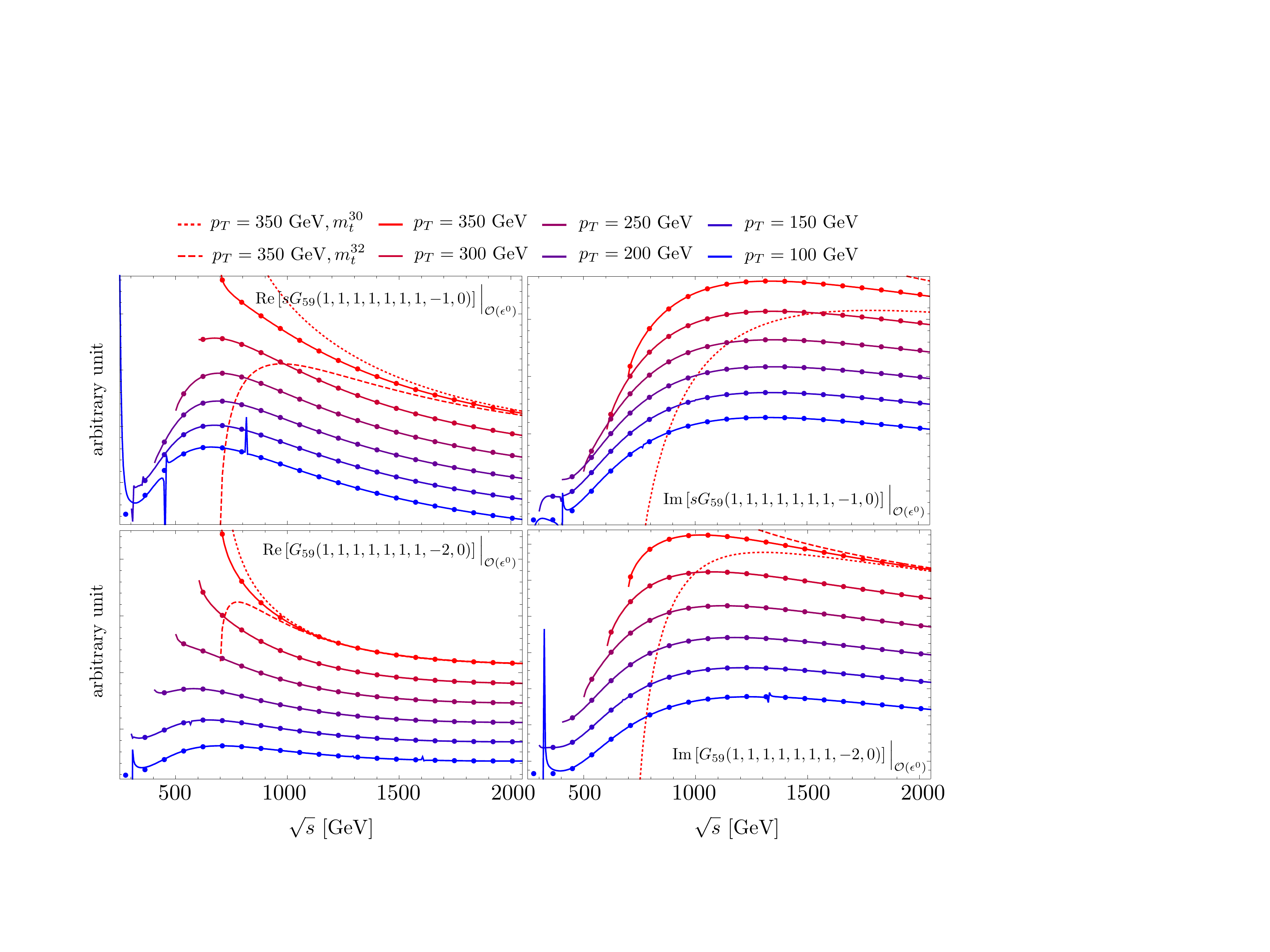}
  \caption{\label{fig::MI_pade}
    Real and imaginary parts of the master integrals
    $G_{59}(1,1,1,1,1,1,1,-1,0)$ and $G_{59}(1,1,1,1,1,1,1,-2,0)$ as a
    function of $\sqrt{s}$ and various fixed values of $p_T$.
    Solid lines are obtained from the Pad\'e-improved expansion in $m_t$.
    The values of $p_T$ decrease from top to bottom.
    The dots are numerical results obtained with {\tt pySecDec}, which have small
    error bars which are not visible in the plot. For the renormalization
    scale $\mu=m_t$ has been chosen.}
\end{figure}

In this section we construct $[8/8]$ Pad\'e approximants (see
Eq.~(\ref{eq::Pade})) at the level of the master integrals, for which
numerical results can be obtained using {\tt FIESTA}~\cite{Smirnov:2015mct}
and {\tt pySecDec}~\cite{Borowka:2017idc}.  In Fig.~\ref{fig::MI_pade} we show
the real and imaginary parts of the non-planar seven-line master integrals
$G_{59}(1,1,1,1,1,1,1,-1,0)$ and $G_{59}(1,1,1,1,1,1,1,-2,0)$ (see
Refs.~\cite{Davies:2018ood,Davies:2018qvx} for the notation and graphical
representations) as a function of $\sqrt{s}$.  In each panel several lines are
shown which correspond to different choices of $p_T$. For better readability
we shift some of the lines such that they are well separated, at least in some
parts of the phase space, which leads to arbitrary units on the $y$-axis.
Solid lines correspond to the Pad\'e approximant\footnote{Similar results are
  also obtained for other choices.} $[8/8]$ and the dots are obtained using
{\tt pySecDec}. One observes an impressive agreement between the
Pad\'e{}-improved and numerical results, even for the lower $p_T$ values
around 100--200~GeV (the lower, blue-coloured lines).  The small spikes
visible above $\sqrt{s} = 500$~GeV in some of the plots are due to the
proximity of poles in the complex plane of the $[8/8]$ Pad\'e approximants. In
our final results, such spikes are removed by the re-weighting procedure
described at the end of Section~\ref{sec::virt}.

For illustration we show for $p_T=350$~GeV the results of the asymptotic
expansions up to order $m_t^{30}$ and $m_t^{32}$ as dashed curves. For
$\sqrt{s}\approx 2000$~GeV reasonable agreement is found with the numerical
result and the Pad\'e approximation.  However, for smaller values of
$\sqrt{s}$ one observes that the expansions quickly deviate from the exact
result.

We obtain similar results for all non-planar master
integrals and are thus confident that the procedure of Section~\ref{sec::virt}
applied to ${\cal V}_{\rm fin}^N$ will provide a good approximation, even for relatively
small values of $p_T$.


\section{\label{sec::Vfin}Numerical results for ${\cal V}_{\rm fin}$}

In this section we consider ${\cal V}_{\rm fin}$ as a function of $\sqrt{s}$ and
$p_T$ and compare to the exact results obtained in~\cite{Borowka:2016ypz}. The
results of~\cite{Borowka:2016ypz} are available from~\cite{hhgrid} in the form of a
grid in the $\sqrt{s}$--$p_T$ plane, where an uncertainty from numerical
integration is assigned to each data point. For the renormalization scale the
value $\mu=\sqrt{s}/2=m_{hh}/2$ has been chosen. Furthermore we use the values
$m_t = 173~\mbox{GeV}$ and $m_h = 125~\mbox{GeV}$.

\begin{figure}[t]
  \centering
  \includegraphics[width=\textwidth]{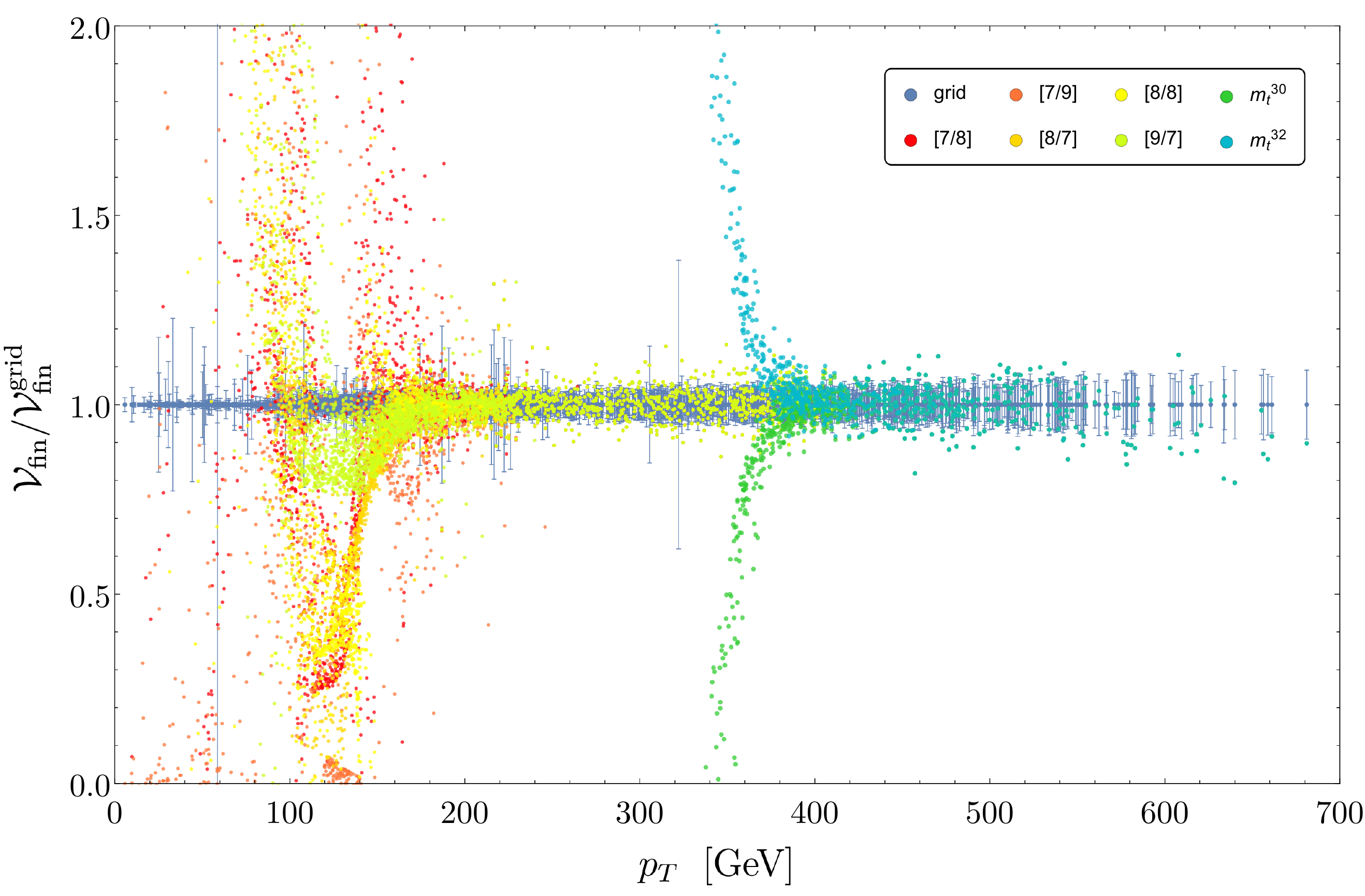}
  \caption{\label{fig::grid} ${\cal V}_{\rm fin}$ normalized to the central
    values provided in~\cite{hhgrid} as a function of $p_T$. The dark blue
    data points with error bars are obtained
    from~\cite{Borowka:2016ypz,hhgrid}.  The data points without uncertainty
    bars are based on ${\cal V}_{\rm fin}^N$, see legend and text for
    details.}
\end{figure}

In Fig.~\ref{fig::grid} we show all data points from~\cite{hhgrid}, normalized
to their central values, as a function of $p_T$ (the dark blue points with
uncertainty bars). Note that in general, different data points belong to
different values of $\sqrt{s}$.  Fig.~\ref{fig::grid} also contains Pad\'e
results for ${\cal V}_{\rm fin}^N$ constructed from $N=30$ and $N=32$ input,
again normalized to the central values of the grid points
from~\cite{hhgrid} (the coloured points without uncertainty bars).
Additionally, the results of the expansions ${\cal V}_{\rm fin}^{30}$
and ${\cal V}_{\rm fin}^{32}$ are shown as green and light-blue data points,
respectively.  Note that the data points based on ${\cal V}_{\rm fin}^N$ are
computed using the same input values as those of the grid points.

As expected, good agreement is found for large values of $p_T$ (which implies
large values of $\sqrt{s}$). Most of the data points lie within one sigma
of the grid points~\cite{hhgrid}. One also observes that some of the
points are outside the one-sigma range, however, still agree within
two sigma. The interesting region of Fig.~\ref{fig::grid} is
$p_T\lesssim400$~GeV. Here, the high-energy expansion diverges rather
quickly and the agreement with the grid points breaks down. The
Pad\'e-improved results, however, follow the dark blue points until
$p_T\approx 200$~GeV. Some of the Pad\'e approximants reproduce the exact
numerical result even down to $p_T\approx 150$~GeV with reasonable
precision. This behaviour motivates a closer look into the
comparison of Pad\'e-improved and numerical results for fixed values
of $p_T$.

We now fix $p_T$ and consider ${\cal V}_{\rm fin}$ as a function of
$\sqrt{s}$. For small values of $p_T$ and $\sqrt{s}$ the grid points are
dense. However, for $p_T\gtrsim 300$~GeV and/or
$\sqrt{s}\gtrsim 1000$~GeV they become quite sparse. Furthermore, if
one wants to perform an analysis for fixed $p_T$ one can in principle only use
a few data points from the grid which makes a comparison difficult.  On the
web-page~\cite{hhgrid} an interpolation routine is provided which allows for an
extension of the grid points to the whole phase space.
However, we find that in regions where the grid is only sparsely populated
this interpolation routine seems to provide unreliable results.  In order to
separate interpolated points with solid support from nearby grid-points from
interpolated points without such support, we enhance an
interpolated data point at $P_0=\left(\sqrt{s}_0,p_{T,0}\right)$
by an error estimate as follows:
\begin{itemize}
\item
  Define a region around $P_0$ as\\ $\Delta =
    \left\{ (\sqrt{s},p_T) \Big{|} 
      |\sqrt{s}-\sqrt{s_0}| \leq 5~\mathrm{GeV},
      |p_{T}-p_{T,0}| \leq 10~\mathrm{GeV}
    \right\}$.
\item $\mathbb{P}$ is the set of data points of the grid~\cite{hhgrid} which
  lie in $\Delta$:
  $\mathbb{P} = \{{\cal V}_{\rm fin}^{\rm grid}(P_1)\pm\delta_1,{\cal V}_{\rm
    fin}^{\rm grid}(P_2)\pm\delta_2,\ldots,{\cal V}_{\rm fin}^{\rm grid}(P_n)\pm\delta_n\}$, where
  $\delta_i$ are the corresponding uncertainties.
\item If $\mathbb{P}$ is empty no uncertainty can be assigned to the
  interpolated value ${\cal V}_{\rm fin}^{\rm int}(P_0)$.  Note that such a
  point has no support from the actual grid points.
\item 
  For non-empty set $\mathbb{P}$ we  define
  $\sigma = \sum_{i=1}^{n}\left|\delta_i\right|/n$ as a mean uncertainty
  assigned to ${\cal V}_{\rm fin}^{\rm int}(P_0)$.
\end{itemize}  

\begin{figure}[t]
  \centering
  \begin{tabular}{cc}
    \includegraphics[width=.48\textwidth]{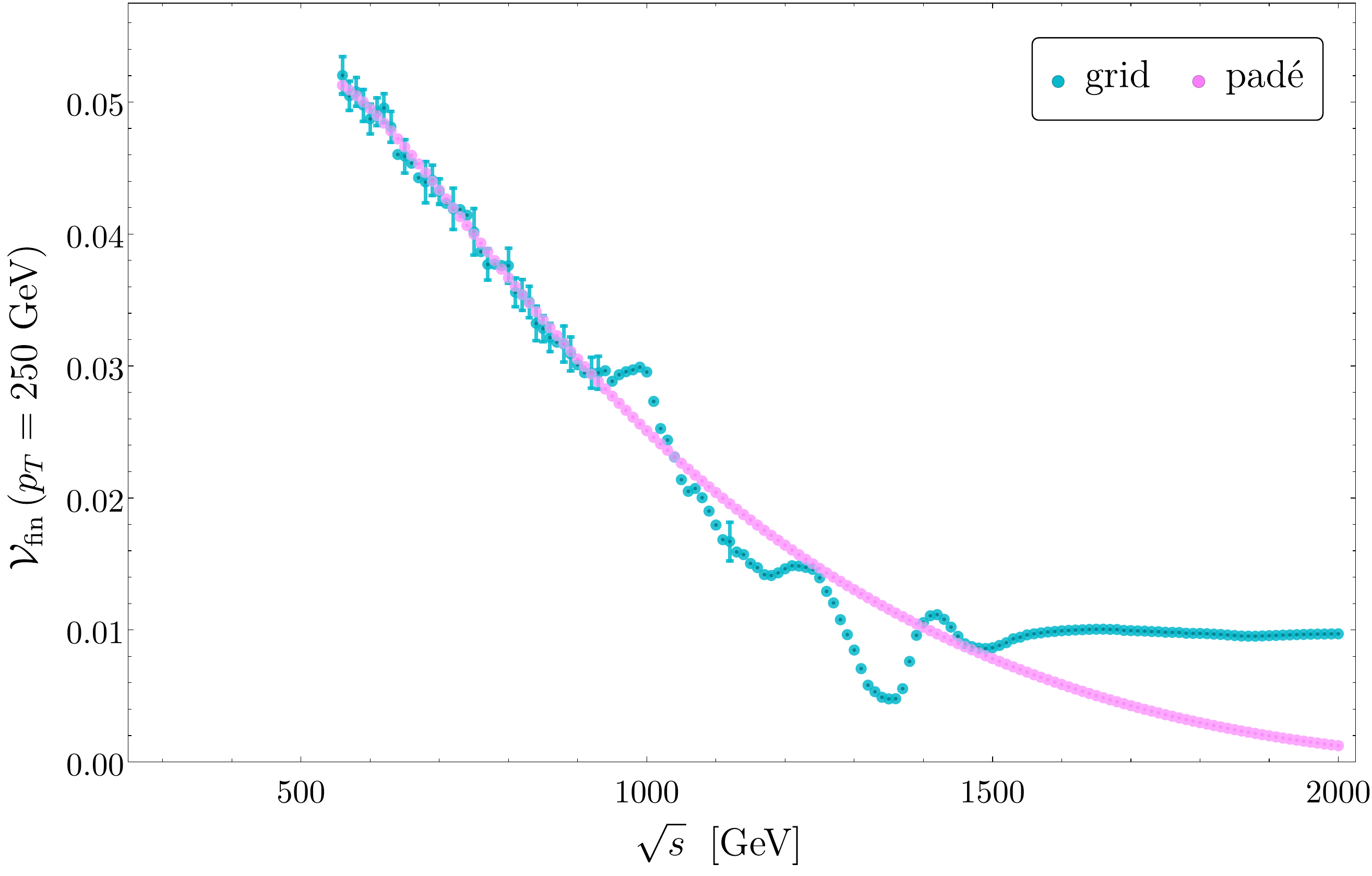} &
    \includegraphics[width=.48\textwidth]{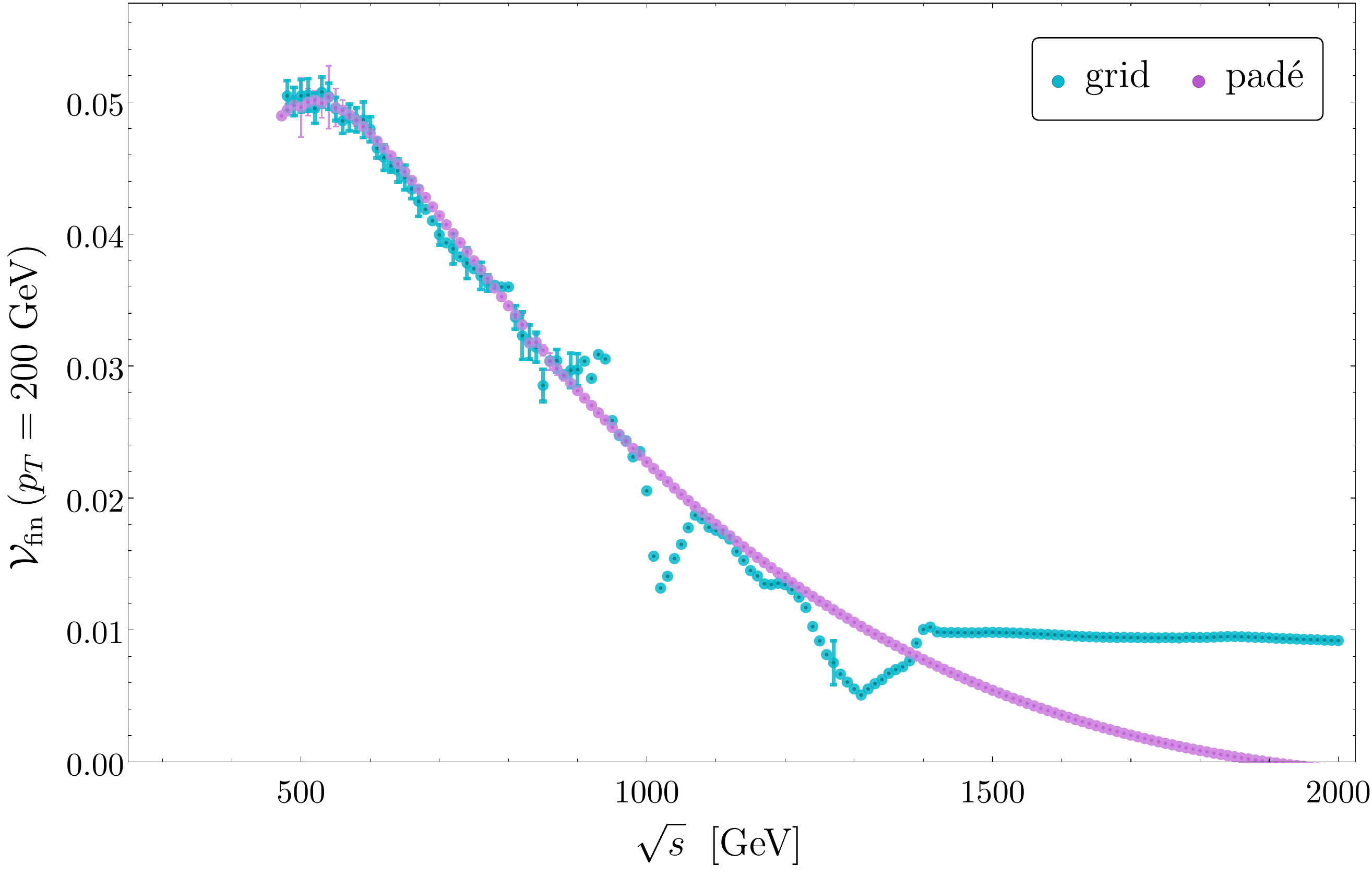} 
    \\ (a) & (b) \\
    \includegraphics[width=.48\textwidth]{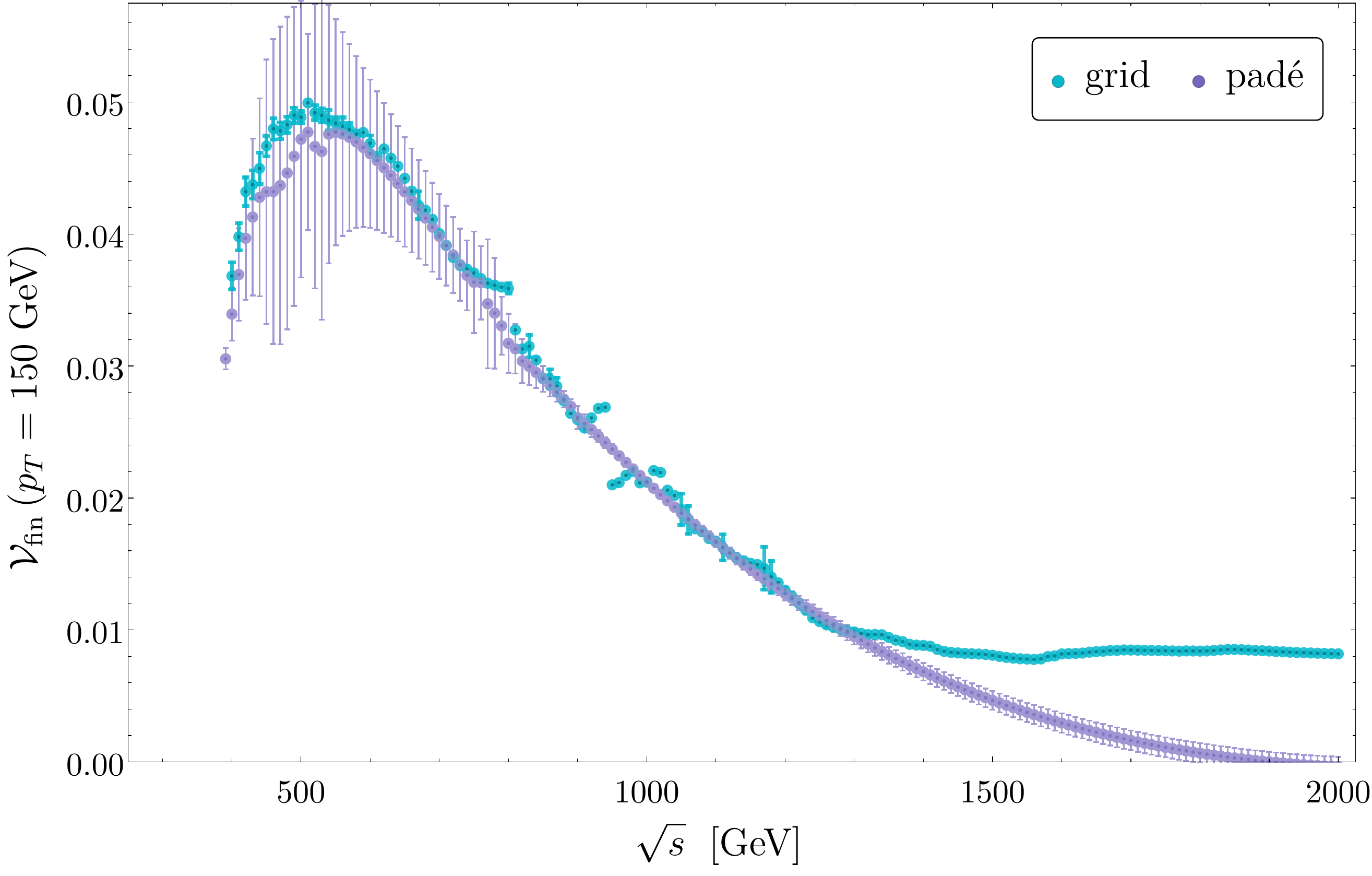} &
    \includegraphics[width=.48\textwidth]{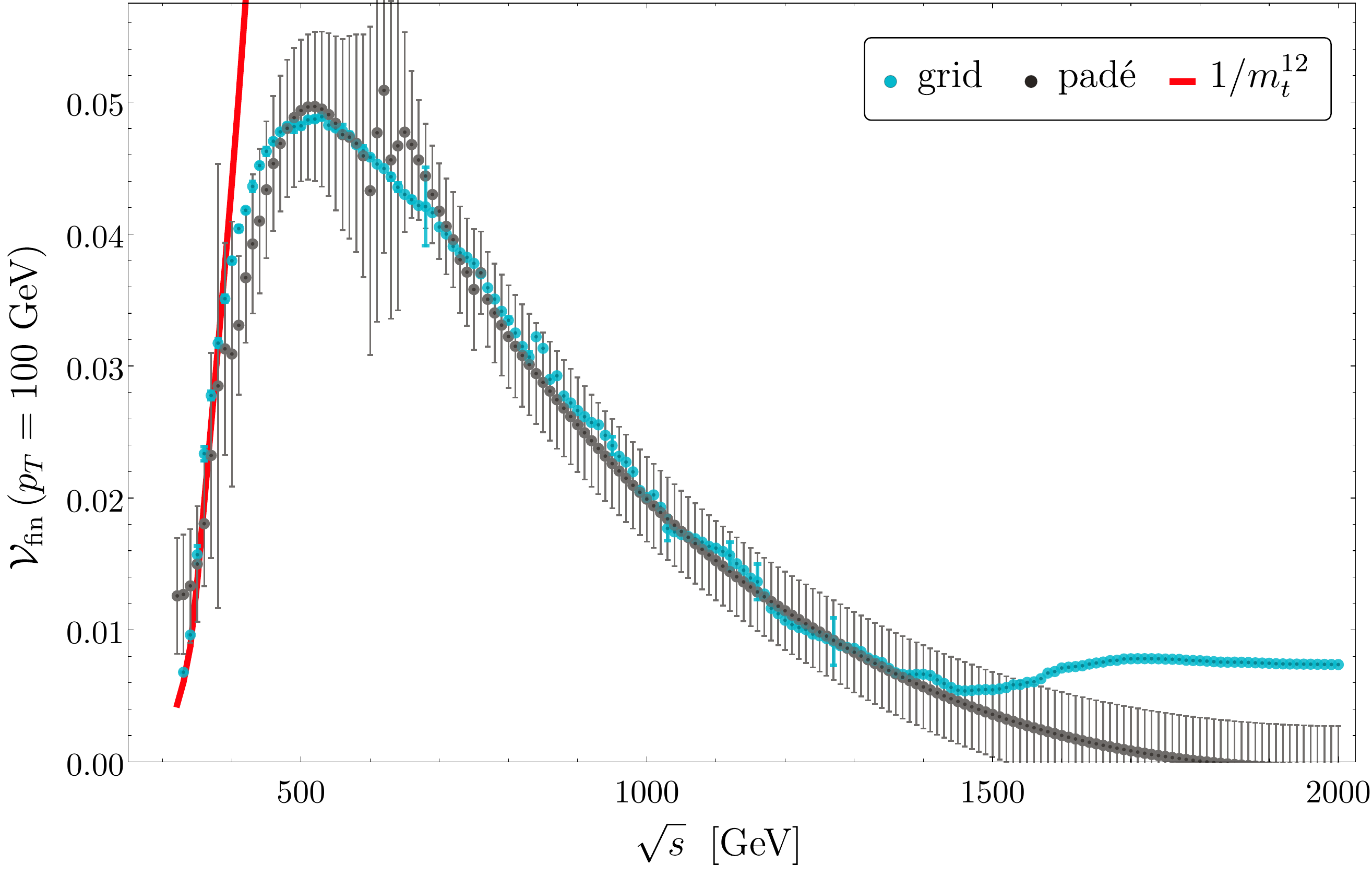}
    \\ (c) & (d)
  \end{tabular}
  \caption{\label{fig::fixed-pt} ${\cal V}_{\rm fin}$ as a function of
    $\sqrt{s}$ for selected values of $p_T$: (a) $p_T=250$~GeV, (b)
    $p_T=200$~GeV, (c) $p_T=150$~GeV, (d) $p_T=100$~GeV.  Both the grid
    points~\cite{Borowka:2016ypz,hhgrid} and the combined Pad\'e improved
    results are shown. For $p_T=100$~GeV we also show the large-$m_t$
    expansion including terms up to order $1/m_t^{12}$.}
\end{figure}

In Fig.~\ref{fig::fixed-pt} we show ${\cal V}_{\rm fin}$ as a function of
$\sqrt{s}$ for four different values of $p_T$. The blue dots correspond to the
results obtained from the grid~\cite{hhgrid} using the procedure described
above. Blue dots with no uncertainty bar have no nearby grid points from
which one can estimate an uncertainty. The other dots correspond to our
Pad\'e-improved results which are obtained using the prescription from
Section~\ref{sec::virt}. If no uncertainty is visible for these points, it
is smaller than the dot size.

Let us start with the discussion of Figs.~\ref{fig::fixed-pt}(a) and~(b) which
correspond to $p_T=250$~GeV and $p_T=200$~GeV, respectively. For $\sqrt{s}<800$~GeV
both the numerical and the Pad\'e results have small uncertainties and agree
very well. Between $\sqrt{s}\approx800$~GeV and $\sqrt{s}\approx1400$~GeV the Pad\'e
results behave smoothly but the (interpolated) numerical results show strong
variation which is due to the interpolation procedure used
in Ref.~\cite{hhgrid}. This is also true for $\sqrt{s}>1400$~GeV where the results
from~\cite{hhgrid} show an unphysical constant behaviour.  This behaviour
suggests that above $\sqrt{s}\approx800$~GeV one should not trust the results
of~\cite{hhgrid} but rather the approximations obtained from the high-energy
expansion~\cite{Davies:2018ood,Davies:2018qvx}.  On the other hand, for
$\sqrt{s}\lesssim800$~GeV, the good agreement of the Pad\'e results with the numerical
calculation provides confidence regarding the reliability of the Pad\'e procedure.

For $p_T=150$~GeV, see Fig.~\ref{fig::fixed-pt}(c), the Pad\'e procedure
develops uncertainties of about 10\% to 20\% for
$\sqrt{s}\lesssim 800$~GeV.  It is nevertheless quite impressive that
agreement with the numerical results, which have small uncertainties, is
found.  For higher values of $\sqrt{s}$ it seems that one can trust the
results from~\cite{hhgrid} up to about $\sqrt{s}=1300$~GeV, above which they
again become constant, which is unphysical.

Although it is far from the region of convergence of the high-energy
expansion, we show in Fig.~\ref{fig::fixed-pt}(d) the results for
$p_T=100$~GeV. Here, the Pad\'e method develops large uncertainties
over the whole range of $\sqrt{s}$.  It is, however, interesting to note
that the central value shows good agreement with the numerical results for
$\sqrt{s}\lesssim1500$~GeV.
In this plot we also show, as a solid red curve, results for the large-$m_t$ expansion of
${\cal V}_{\rm fin}$, which is constructed using the large-$m_t$ expansion of the form
factors, computed to order $1/m_t^{12}$ in~\cite{Grigo:2014jma}.
We observe agreement with the exact results
(blue dots) up to $\sqrt{s}\approx 400$~GeV which constitutes a good
consistency check.

\begin{figure}[t]
  \centering
  \includegraphics[width=\textwidth]{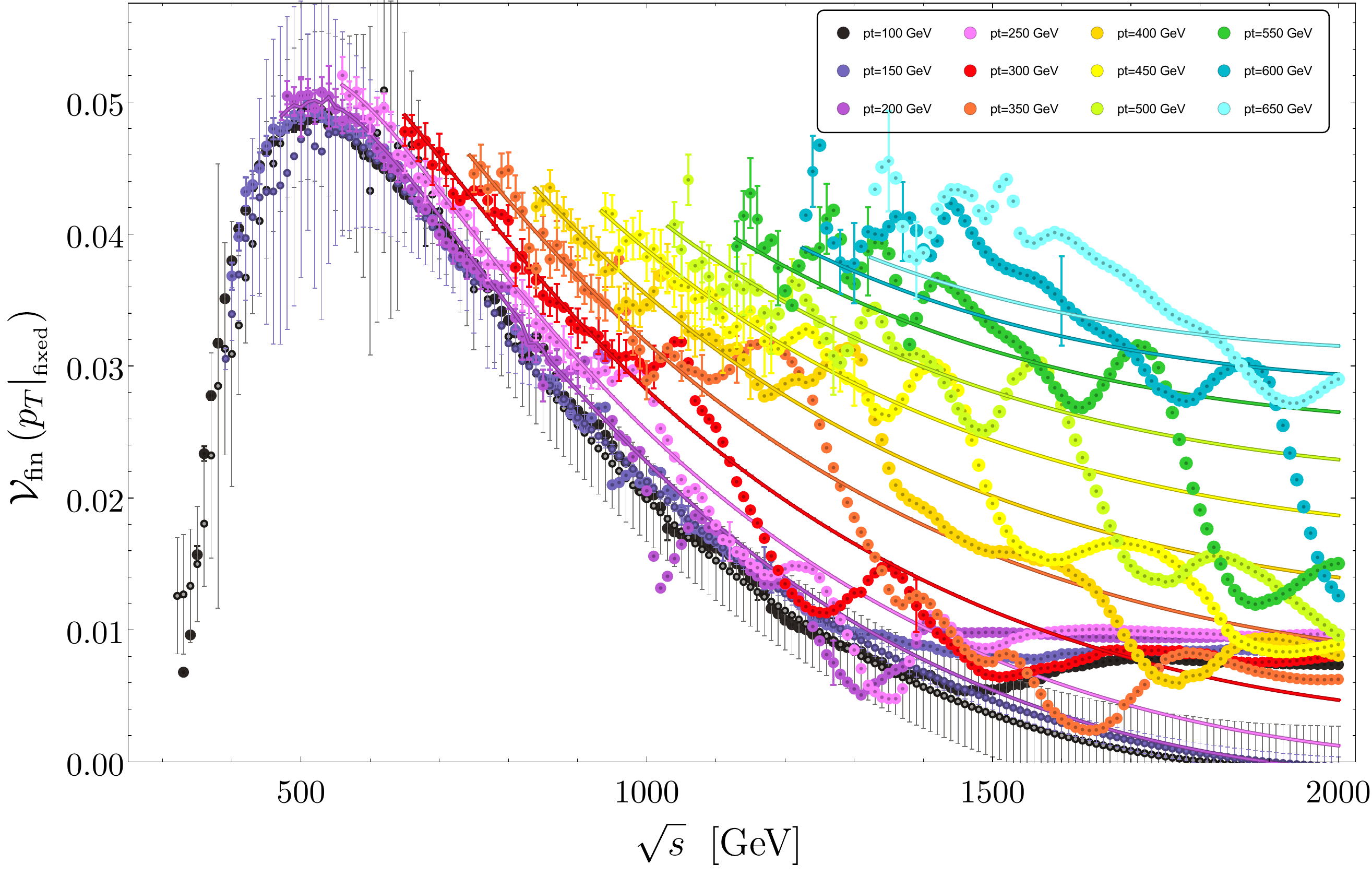}
  \caption{\label{fig::Vfin_pt} ${\cal V}_{\rm fin}$ as a function of
    $\sqrt{s}$ for fixed values of $p_T$. The data points and the
    corresponding uncertainty bars are obtained from the grid~\cite{hhgrid}.
    The solid lines are based on the Pad\'e-improved high-energy
    expansion. For high values of $p_T$ the uncertainties of the Pad\'e results are smaller than the
    thickness of the lines. The uncertainties shown for 
    $p_T$ values below 200 GeV are taken over from Fig.~\ref{fig::fixed-pt}.}
\end{figure}

The discussion of the plots in Fig.~\ref{fig::fixed-pt} shows that the Pad\'e
method provides accurate results even for relatively small values of
$p_T$. Furthermore, it provides realistic estimates of the uncertainties.  In
Fig.~\ref{fig::Vfin_pt} we show ${\cal V}_{\rm fin}$ as a function of
$\sqrt{s}$ for fixed values of $p_T$ (shown in different colours, see the plot
legend for details). The plot contains the curves for the four $p_T$ values of
Fig.~\ref{fig::fixed-pt} and a further eight choices of $p_T$, with the highest
value $p_T=650$~GeV. The dots represent the results from~\cite{hhgrid}. Where
available, the uncertainties are explicitly indicated. For $p_T\ge200$~GeV the
Pad\'e results are shown as solid lines. Note that in this region of the phase
space the uncertainty is below the thickness of the lines.  One observes that
the solid lines agree with the data points within the indicated
uncertainties, which are in general much larger than the Pad\'e uncertainty.
For $p_T=100$~GeV and $p_T=150$~GeV we reproduce in Fig.~\ref{fig::Vfin_pt}
the curves from Fig.~\ref{fig::fixed-pt} (see black and dark violet data points).

\begin{figure}[t]
  \centering
  \includegraphics[width=\textwidth]{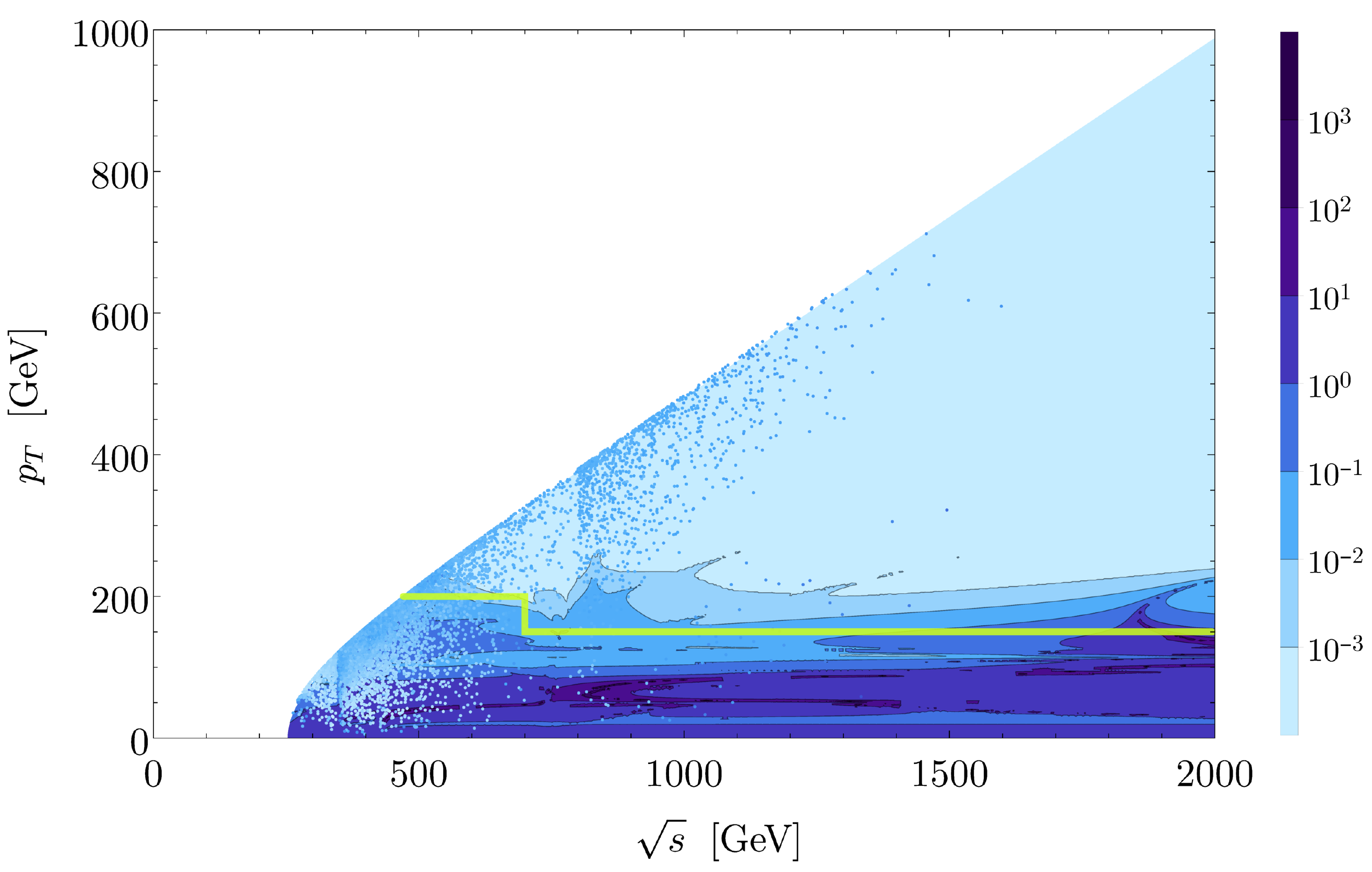}
  \caption{\label{fig::rel_uncert} Relative uncertainty of the Pad\'e results
    in the $\sqrt{s}$--$p_T$ plane. The points of~\cite{hhgrid} are overlayed.
    Note that a logarithmic scale is used for the relative uncertainty.}
\end{figure}

We now define a criterion which provides a prescription
for the improvement of the grid~\cite{hhgrid}. In order to have guidance we
show in Fig.~\ref{fig::rel_uncert} the relative uncertainty of the Pad\'e
results in the $\sqrt{s}$--$p_T$ plane. We also overlay all grid points
from~\cite{hhgrid} and use the same colour scale for their uncertainties.
Note that the kinematic boundary is obtained from the requirement that
$1-4(m_h^2+p_T^2)/s$ (see Eq.~(\ref{eq::t})) is positive.

From Fig.~\ref{fig::rel_uncert} we learn that the uncertainty is below $0.1\%$
for $p_T\gtrsim 200$~GeV and then grows towards lower $p_T$ relatively
quickly.  Still, even for $p_T\approx 150$~GeV the uncertainty is around a few
percent for most values of $\sqrt{s}$. Note that larger relative
uncertainties for larger values of $\sqrt{s}$ are observed since in this
region $\widetilde{\mathcal{V}}_{\textnormal{fin}}$ is small.

On the basis of this observation we extend the
grid provided in~\cite{hhgrid} as follows:
\begin{itemize}
\item We increase the number of points computed using the full NLO result from 3398 to 6320. The new points are 
sampled according to the distribution of unweighted events and, therefore, populate the same kinematic regime as the original points.
\item For $\sqrt{s}\geq 700$~GeV and $p_T\ge 150$~GeV we add points from the Pad\'e approximation.
\item For $\sqrt{s}<    700$~GeV and $p_T\ge 200$~GeV we add points from the Pad\'e approximation.
\end{itemize}
The boundary above which we include points from the Pad\'e approximation is denoted
as a yellow line in Fig.~\ref{fig::rel_uncert}. We note here that if one reproduces
Figs.~\ref{fig::Vfin_pt} and \ref{fig::rel_uncert} using the 6320 points described above the
behaviour is qualitatively the same and we therefore refrain from showing them in this paper.

\begin{figure}[t]
  \centering
  \includegraphics[width=\textwidth]{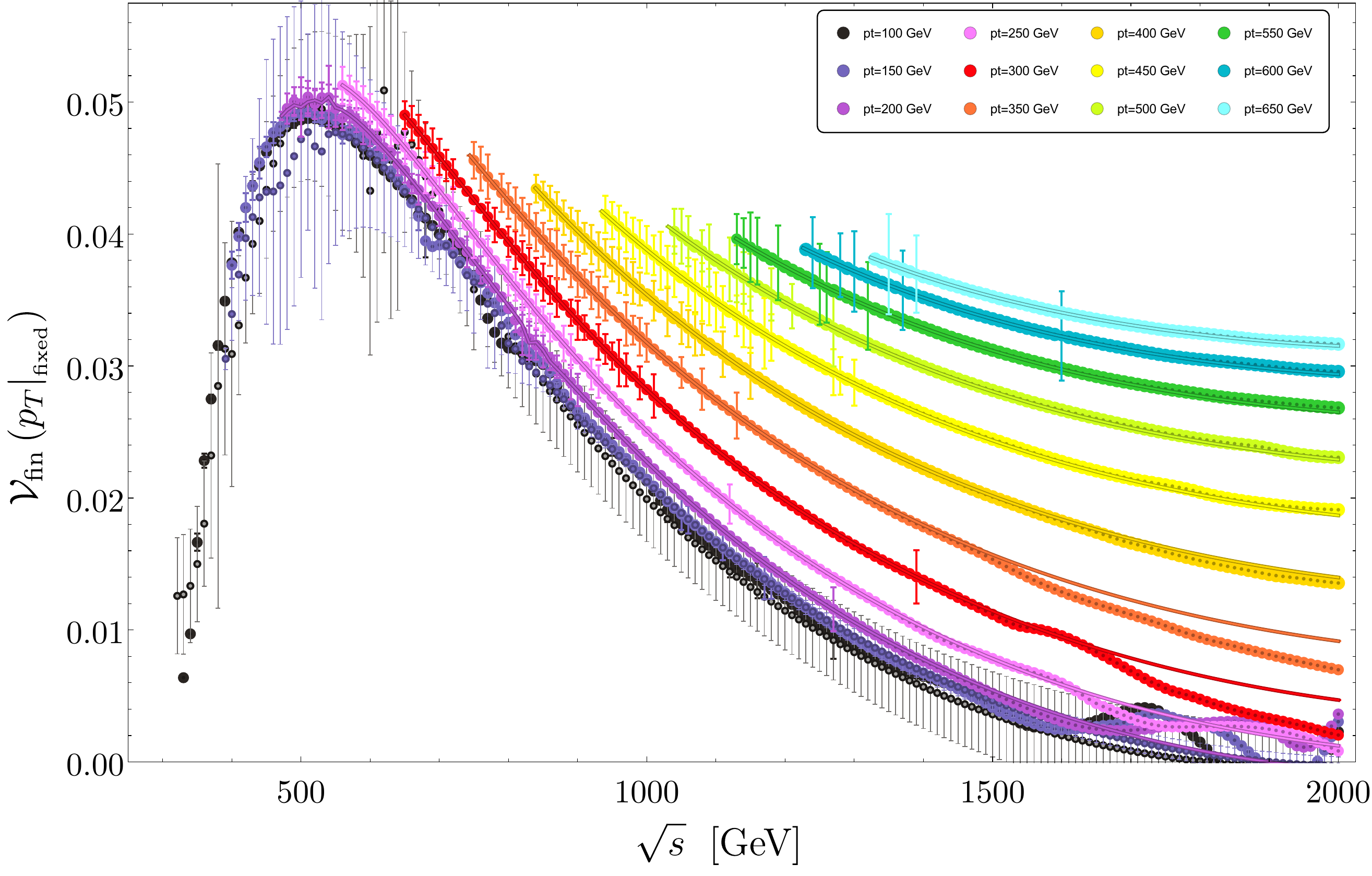}
  \caption{\label{fig::Vfin_pt_impr} ${\cal V}_{\rm fin}$ as a function of
    $\sqrt{s}$ for fixed values of $p_T$. The data points are obtained from
    the improved version of the grid and the solid lines are based on the
    Pad\'e-improved high-energy expansion.  }
\end{figure}

In Fig.~\ref{fig::Vfin_pt_impr} we compare the Pad\'e results to the improved
version of the grid, which provides precise results in the whole relevant
phase space.
We note that the wiggly behaviour and the deviation of the grid data points from
the Pad\'e approximation for larger values of $\sqrt{s}$ and smaller values of $p_T$
could be improved by including further data points from the Pad\'e approximation. This
behaviour would then be pushed to higher values of $\sqrt{s}$. We judge the performance
of the grid as displayed by Fig.~\ref{fig::Vfin_pt_impr} to be sufficient for the
phenomenological applications of this paper, and further improvements of the grid not
to be necessary.
This improved grid can be downloaded from~\cite{hhgrid}.

\section{\label{sec::appl}Applications}

In the following we discuss differential distributions w.r.t.  the Higgs boson pair
invariant mass $m_{hh}$ and the ``single inclusive'' Higgs boson transverse
momentum $p_{T,h}$ for hadronic centre-of-mass energies $\sqrt{s_H}=14$~TeV
and $\sqrt{s_H}=100$~TeV. The emphasis of this analysis is the comparison of
the current~\cite{hhgrid} and improved grid introduced in the previous
section.

For our analysis we use the parton distribution functions
\verb|PDF4LHC15_nlo_100_pdfas|~\cite{Harland-Lang:2014zoa,Dulat:2015mca,Butterworth:2015oua} and adopt
the corresponding value for $\alpha_s$.
For the top quark and Higgs boson masses we use $m_h=125$~GeV and $m_t=173$~GeV and
choose $\mu_0=m_{hh}/2$ as the central value for the renormalization ($\mu_R$)
and factorization ($\mu_F$) scales. The uncertainties due to higher-order QCD corrections
are estimated using the usual seven-point scale
variation around $\mu_0$, i.e., for $\mu_R$ and $\mu_F$ we introduce
$\mu_{R,F}=c_{R,F}\mu_0$ with $c_{R,F}\in\{0.5,1,2\}$ and omit the extreme
choices $(c_R,c_F)=(0.5,2)$ and $(c_R,c_F)=(2,0.5)$.

\begin{figure}[t]
  \centering
  \mbox{
    \includegraphics[width=.5\textwidth]{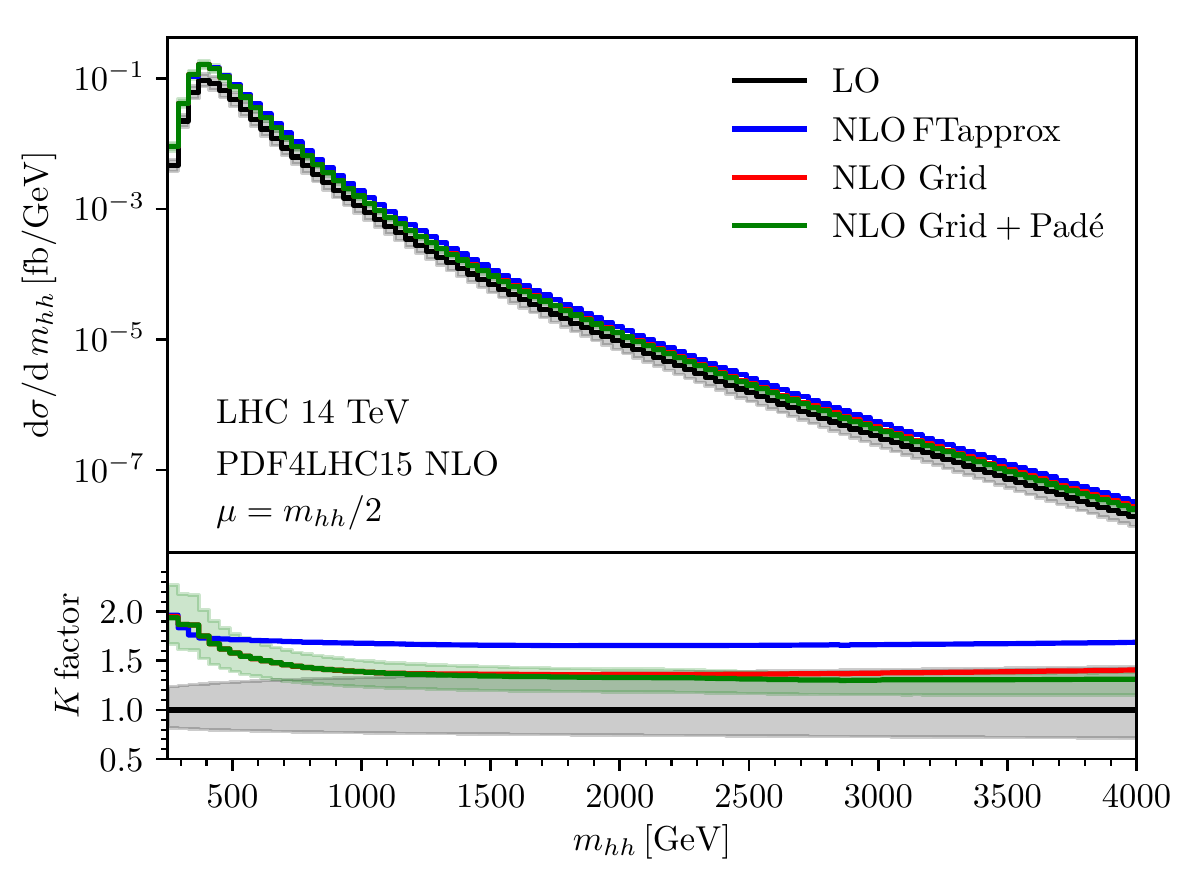}
    \hfill
    \includegraphics[width=.5\textwidth]{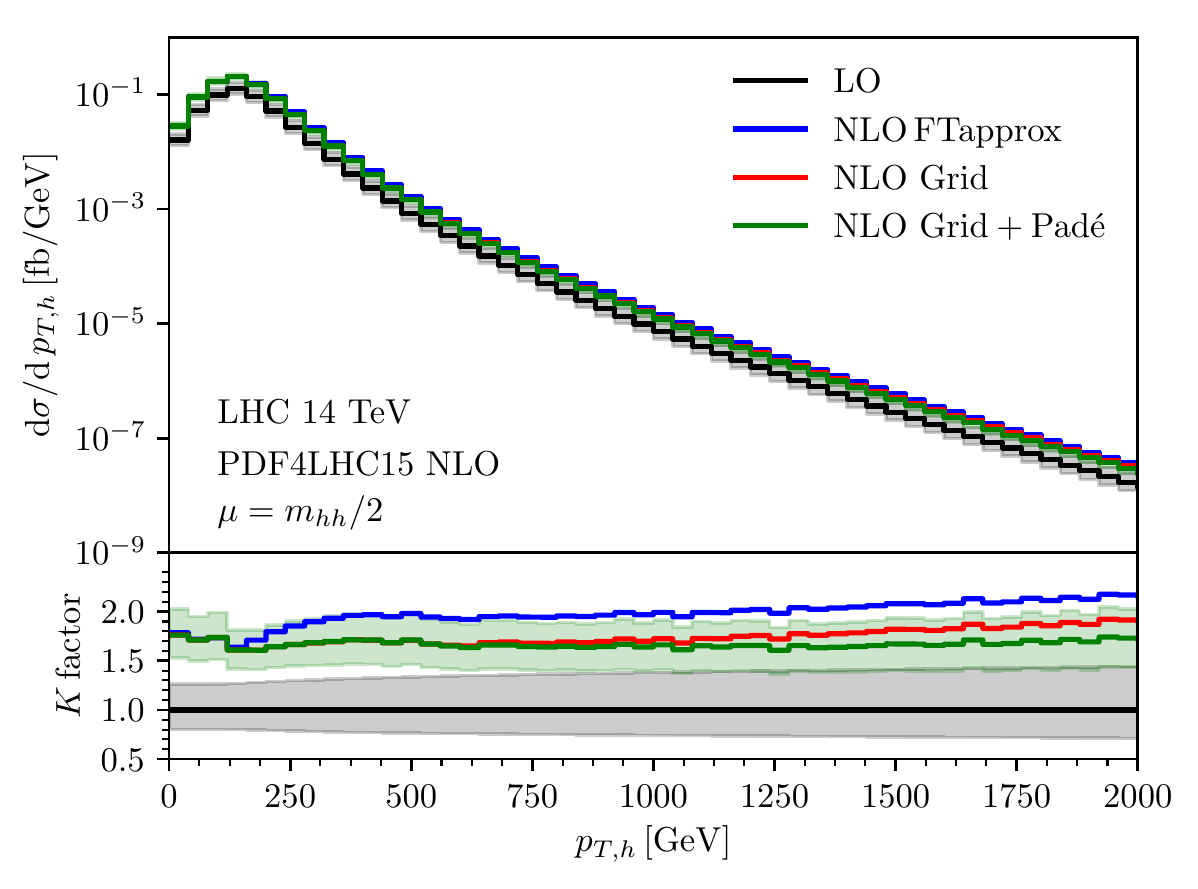}
  }
  \caption{\label{fig::mhhpt14cut}$m_{hh}$ and $p_{T,h}$ distributions for a
      hadronic centre-of-mass energy $\sqrt{s_H}=14$~TeV.}
\end{figure}

In Fig.~\ref{fig::mhhpt14cut} we show our results for $\sqrt{s_H}=14$~TeV.  In
the upper panels we present the $m_{hh}$ and $p_{T,h}$ differential distributions, and in
the lower panels we display the ratio of the NLO corrections to the LO values
($K$ factor). The LO values are shown in black and the coloured curves
correspond to different versions of the NLO prediction,  all of which contain
the full real radiation corrections and only differ in the way that the virtual
corrections are implemented.  The blue curve, denoted ``FTapprox'',
incorporates the virtual corrections computed in the infinite top quark mass
limit and rescaled by the exact LO prediction. The red curve is based on the
grid constructed in Ref.~\cite{Heinrich:2017kxx} but improved by increasing the
number of points from 3398 to 6320 (see discussion above).  Finally, the green
curve is based on the new grid, the construction of which is described in
Section~\ref{sec::Vfin}.
This curve constitutes our best prediction. The grey and green
bands around the corresponding curves have been obtained by independent
variations of $\mu_R$ and $\mu_F$ as described above.

It is interesting to note that for small $m_{hh}$ and $p_{T,h}$ there is perfect
agreement of the red and green curves, which is expected since in this region the
dependence on ${\cal V}_{\rm fin}$ comes primarily from the region in the (partonic)
$\sqrt{s}$--$p_T$ plane where the support of the old grid was dense.
For higher values of $m_{hh}$ and $p_{T,h}$, one observes a difference between
the red and the green curves. However, in both cases the red curve lies well
within the green uncertainty band.

\begin{figure}[t]
  \centering
  \mbox{
    \includegraphics[width=.5\textwidth]{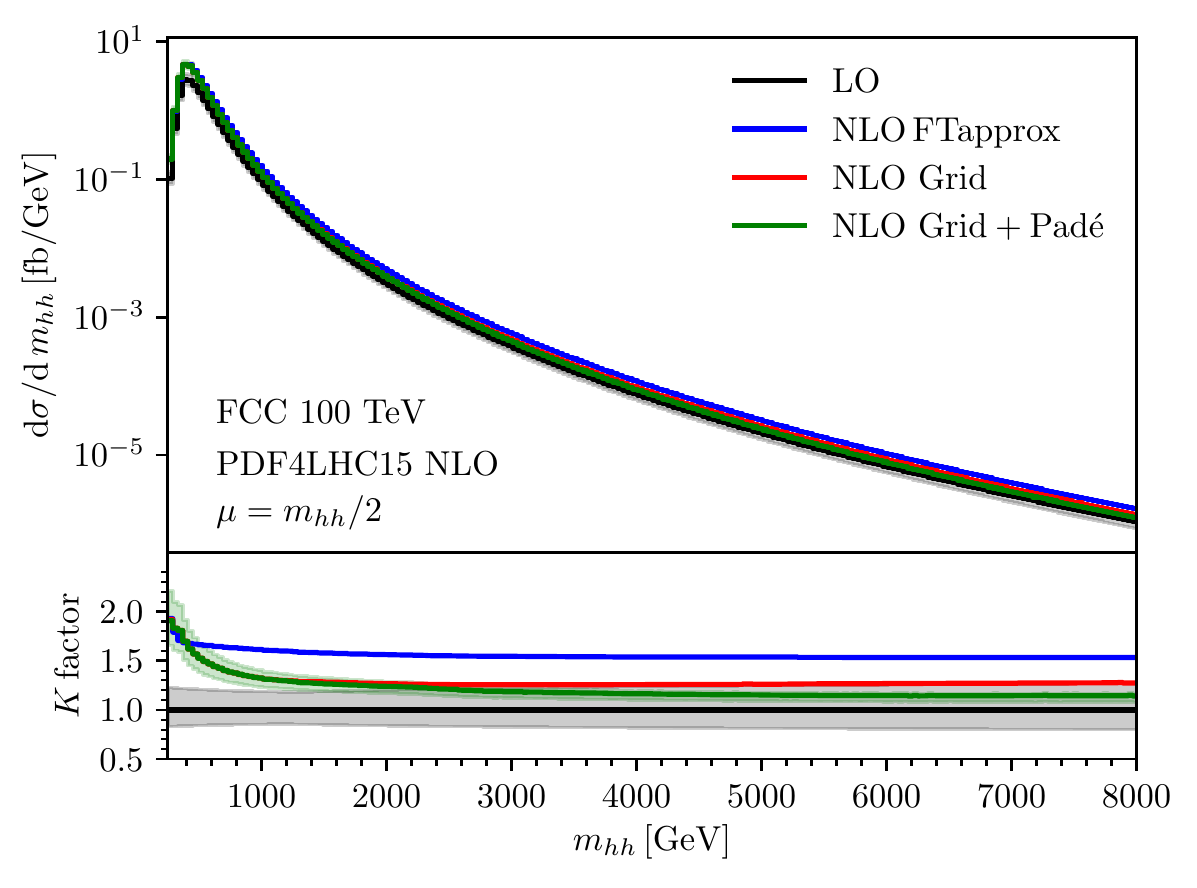}
    \hfill
    \includegraphics[width=.5\textwidth]{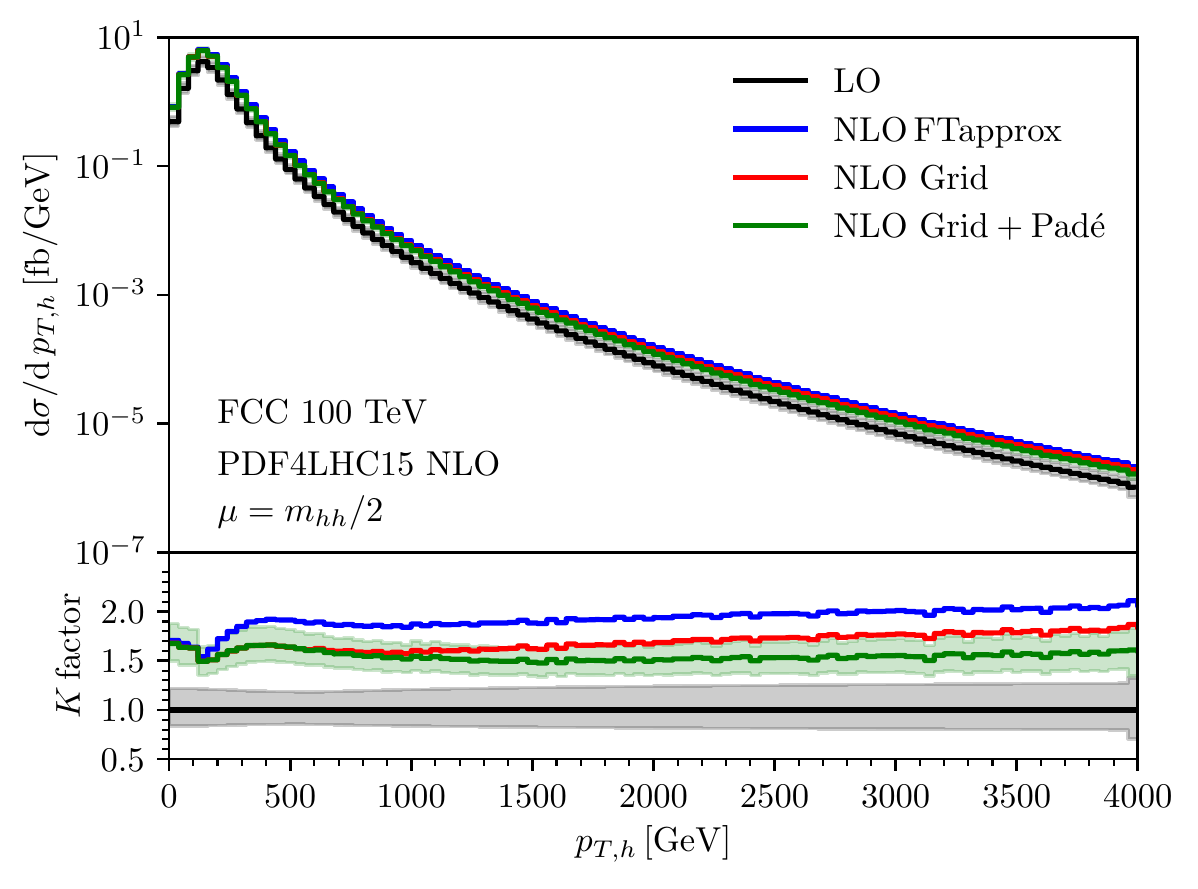}
  }
  \caption{\label{fig::mhhpt100cut}$m_{hh}$ and $p_{T,h}$ distributions for a
    hadronic centre-of-mass energy $\sqrt{s_H}=100$~TeV.}
\end{figure}

The $m_{hh}$ and $p_{T,h}$ distributions for $\sqrt{s}_H=100$~TeV
are shown in Fig.~\ref{fig::mhhpt100cut}, where the same notation is used as
in Fig.~\ref{fig::mhhpt14cut}. Note that now a significant difference is
observed between the red and green curves; for higher values of $m_{hh}$ and
$p_{T,h}$ the red curve lies outside the green uncertainty band.
As an example let us consider $p_{T,h}=2000$~GeV. For this value
the $K$ factor is reduced from $K\approx1.7$ to
$K\approx1.5$ after including the high-energy results
in the grid. 

Let us mention that in Figs.~\ref{fig::mhhpt14cut} and~\ref{fig::mhhpt100cut},
the same phase-space points have been used for all curves. Thus, the
differences between the curves is only due to the different implementations of the
virtual corrections.

We should emphasize that one observes no change in the total cross section due
to the change from the red to the green curve, since the main contribution to
$\sigma_{\rm tot}$ comes from smaller centre-of-mass energies.  However,
Figs.~\ref{fig::mhhpt14cut} and~\ref{fig::mhhpt100cut} show that it is
important to use the improved grid for phenomenological analyses, if
one wishes to consider large values of $m_{hh}$ or $p_{T,h}$, even for
$\sqrt{s_H}=14$~TeV. In these regions the predictions based on
``FTapprox'' deviate significantly from the green curve.


\section{\label{sec::concl}Conclusions}

We provide optimized predictions for the NLO corrections to
Higgs boson pair production by combining the exact numerical results
with analytic expressions for the form factors obtained in a
high-energy expansion. For the latter the region of convergence is
significantly improved by constructing Pad\'e approximants, which are
validated at the level of master integrals. Furthermore, we identify
regions in the phase space where both the exact numerical evaluations
and the Pad\'e results provide precise predictions and find good
agreement. We thus combine both approaches and generate a new grid
which is available from~\cite{hhgrid}. The analytic expressions for the
high-energy expansion of the form factors are available
from~\cite{progdata_appl}.

We apply the improved grid to phenomenological studies of the Higgs boson 
pair invariant mass and Higgs boson transverse momentum distributions at LHC
energies and for $\sqrt{s_H}=100$~TeV.  We show that at high energies
the improvements are noticeable and we recommend to use the updated grid
for phenomenological studies, even for $\sqrt{s_H}=14$~TeV.

 
\section*{Acknowledgements}

This research was supported in part by the COST Action CA16201
(`Particleface') of the European Union and by the Deutsche
Forschungsgemeinschaft (DFG, German Research Foundation) under grant 396021762
--- TRR 257 ``Particle Physics Phenomenology after the Higgs Discovery''.  MK
was supported in part by the Swiss National Science Foundation (SNF) under
grant number 200020-175595.  DW acknowledges the support of the DFG-funded
Doctoral School KSETA.



\end{document}